\documentclass{emulateapj}

\usepackage{amsmath}
\usepackage{xcolor}
\usepackage{ulem}
\usepackage{mathrsfs}
\usepackage{float}
\usepackage[caption=false]{subfig}

\newcommand{\EMRICode}[1]{\texttt{EMRICode#1 }}
\DeclareMathOperator{\Tr}{Tr}

\begin{document}

\title{A Multipole Technique for the Simulation and Analysis of Gravitational Waveforms from Extreme Mass Ratio Inspirals}

\author{Courtney~A.~McIntosh\altaffilmark{1,2}}
\author{Jeremy~D.~Schnittman\altaffilmark{2}}
\email{courtney.mcintosh@bison.howard.edu}

\altaffiltext{1}{Department of Physics and Astronomy, Howard University, Washington, DC 20059, USA}
\altaffiltext{2}{Code 661, Multimessenger Astrophysics Branch, NASA Goddard Space Flight Center, Greenbelt, MD 20771, USA}

\begin{abstract}
We introduce \EMRICode{,}a Python-based numerical framework for evolving generic extreme mass-ratio inspirals (EMRIs) in Kerr spacetime. The code combines adaptive numerical integration of test-particle geodesics with a prescribed model for the secular evolution of the orbital constants of motion, allowing eccentric and inclined trajectories to be followed over long inspiral timescales. We focus in particular on the treatment of the Carter constant, whose evolution is not directly determined by the leading-order fluxes for energy and axial angular momentum used in the present model. We investigate several prescriptions for determining its evolution and find that a radial-potential-based prescription provides the most stable behavior among those considered, while also exhibiting a systematic residual bias for nearly planar orbits. We compare the resulting inspirals with established radiation-reaction approximations, including the weak-field treatment of Peters and Kerr-based prescriptions from the literature, and use these comparisons to assess the qualitative behavior and limitations of the implementation. The modular structure of \EMRICode{}allows the radiation-reaction prescription to be replaced independently of the underlying geodesic evolution, providing a framework for incorporating more accurate flux models and additional physical effects. We present two simple examples of how interactions with accretion disks and partial tidal disruptions can drive the evolution of a stellar EMRI orbit. \EMRICode{}therefore provides a flexible computational framework for studying the orbital dynamics of EMRIs and for exploring extensions beyond the idealized vacuum, point-particle approximation.
\end{abstract}

\section{Introduction}
\label{sec:intro}
Extreme Mass Ratio Inspirals (EMRIs) are systems in which a compact object, such as a stellar-mass black hole or neutron star, orbits a supermassive black hole (SMBH) with a mass ratio $\mu/M \lesssim 10^{-4}$. In these systems, the primary driver of orbital evolution is the emission of gravitational radiation \citep{AmaroSeoane2007}. EMRIs are of considerable astrophysical interest due to their potential as precise probes of SMBH properties. The gravitational waveforms emitted during an inspiral encode detailed information about the mass and spin of the central object, offering insights into its formation history and the dynamical environment of galactic nuclei. Because the compact object orbits deep within the gravitational potential of the SMBH, EMRIs serve as excellent testbeds for general relativity in the strong-field regime.

With the upcoming Laser Interferometer Space Antenna (LISA) mission, the need for accurate and computationally efficient EMRI waveform models is becoming increasingly important. The long duration and rich harmonic structure of EMRI signals make them particularly sensitive to small errors in the orbital evolution, while the large range of possible orbital configurations creates a substantial computational challenge for waveform generation and parameter-space exploration. 

A variety of approaches have therefore been developed to model EMRIs, with different methods making different compromises between physical fidelity and computational cost. At one end of this range are black-hole perturbation theory and gravitational self-force approaches, which treat the compact object as a perturbation of the background spacetime and provide a systematic framework for calculating its gravitational backreaction \citep{BarackPound2019}. In Kerr spacetime, the Teukolsky formalism provides the basis for calculating gravitational radiation from the perturbed system, with frequency-domain methods allowing the perturbation equations to be separated into radial and angular components \citep{Teukolsky1973}. Teukolsky-based calculations have subsequently been applied to generic eccentric and inclined geodesics, demonstrating the rich harmonic structure of the resulting waveforms \citep{DrascoHughes2006,FujitaHikidaTagoshi2009}. Time-domain approaches have likewise been extended from simple circular equatorial trajectories to dynamically evolving generic orbits, including approximate inspiral trajectories driven by radiation-reaction prescriptions \citep{Sundararajan_et_al_2008}. Although these methods can be implemented efficiently for specific configurations, the high-fidelity description of dynamics and emitted radiation generally requires a large number of harmonic modes for generic orbits, and the computational cost of evaluating so many amplitudes can make extensive parameter-space exploration challenging. 

A complementary class of approaches seeks to reduce this computational cost by exploiting the separation between the orbital and inspiral timescales. On the orbital timescale, the compact object can be treated as moving along an approximately fixed Kerr geodesic, while the orbital constants evolve on the much longer inspiral timescale due to radiation reaction; adiabatic and flux-driven models take advantage of this structure by evolving these quantities using orbit-averaged fluxes or approximate prescriptions \citep{Babak_et_al_2007, BarackPound2019, LISAConsort2025}.

"Kludge" models extend this philosophy further by combining approximate descriptions of orbital evolution and waveform generation. Analytic kludges use simplified analytic descriptions of the orbital motion and gravitational-wave emission, as in the model of \citet{BarackCutler2004}, while numerical kludges retain a relativistic treatment of the orbital trajectory but use an approximate prescription for the waveform generation \citep{Babak_et_al_2007}. These approaches can be generated substantially faster than full perturbative calculations, making them useful for broad parameter-space studies and the development of large waveform banks, but the approximations introduced into the orbital dynamics or waveform generation can limit their accuracy in the strong-field regime. 

More recent approaches have sought to retain more sophisticated relativistic calculations while reducing their computational cost through techniques such as interpolation, reduced-order representations, and GPU acceleration\citep{Chua2021, Speri_et_al_2024, LISAConsort2025}. Machine-learning methods have been applied to EMRI waveform generation and downstream tasks such as signal detection and parameter estimation, although these approaches generally address the computational cost of waveform evaluation or data analysis rather than providing a general-purpose model of the underlying orbital dynamics. For example, FastEMRIWaveforms (FEW) combines separately constructed trajectory, flux, and waveform modules to enable rapid generation of relativistic EMRI waveforms \citep{Chua2021,Katz2021}. FEW provides sophisticated models that rapidly evaluate inspiral trajectories, including a generic Kerr model based on an augmented analytic kludge and more specialized ODE models for eccentric Schwarzschild and eccentric equatorial Kerr systems, but, these models remain subject to the limitations and parameter ranges of their underlying approximations and precomputed interpolations. In particular, the accuracy of post-Newtonian expansions used in the generic Kerr model are known to degrade in the strong-field regime \citep{Sago_Fujita_2015, Fujita_Shibata_2020}, while the more specialized models are necessarily restricted to the regions of parameter space for which suitable precomputed data can be generated. As such, while this approach provides substantial computational advantages, broad coverage of the EMRI parameter space remains a significant challenge.

In this work we present \EMRICode{,}a numerical integrator designed to model the evolution of EMRIs by solving geodesic motion in Kerr spacetime with radiation-reaction corrections. Rather than replacing the relativistic orbital dynamics with a fully analytic approximation, \EMRICode{}evolves the geodesic trajectory directly while treating radiation reaction as a separate, modular component. This approach retains the generic orbital structure of Kerr geodesics while allowing the radiation-reaction prescription to be modified independently. The resulting framework is intended to provide a balance between relativistic fidelity and computational efficiency, enabling rapid exploration of a broad range of orbital configurations without requiring the computational cost of a full self-force calculation. 

The modular treatment of the non-geodesic forces also provides a natural framework for extending the model beyond gravitational radiation reaction. For example, additional perturbations associated with interactions between the compact object and an accretion disk, or with tidal disruption and related environmental effects, can be incorporated as additional contributions to the evolution of the orbital constants in a straightforward manner. These extensions are beyond the scope of the present work, but motivate the development of a flexible framework in which the underlying geodesic evolution and the mechanisms driving secular orbital evolution can be treated separately.

\section{Design Goals}
\label{Design Goals}

The development of \EMRICode{}was guided by a set of core design principles intended to balance physical fidelity, computational efficiency, and usability. Although the code is capable of producing detailed inspiral trajectories and waveforms, its primary purpose is to serve as a broadly applicable, flexible tool for exploring EMRI dynamics across a wide range of configurations. Several goals shaped this design:

\begin{itemize}

\item \textbf{Broad Parameter Coverage.}  
EMRI systems depend on a large number of orbital and physical parameters, many of which introduce significant computational complexity. As a result, many existing approaches focus on restricted subsets of parameter space, such as equatorial, quasi-circular, or low-spin configurations to simplify calculations. While efficient, these restrictions limit the generality of the resulting datasets. In contrast, \EMRICode{}is designed to evolve inspirals across as broad a region of parameter space as possible, allowing users to probe regimes that are often neglected due to their computational cost or analytic difficulty. 

\item \textbf{Utility as a Preliminary Exploration Tool.}  
Broad parameter coverage inevitably comes with trade-offs: a tool capable of spanning a broad range of EMRI configurations may not match the precision or specialization of methods tailored to specific orbital families. The aim of \EMRICode{}is therefore not to replace high-accuracy, domain-specific solvers, but to complement them. The code should reliably capture qualitative and semi-quantitative behavior across diverse configurations so that results can be compared with more specialized self-force or Teukolsky-based tools when those methods are applicable to the same configuration. It is intended as a fast and flexible first pass that can guide more detailed follow-up work.

\item \textbf{Ease of Use and Transparency.} 
Although the underlying physics and numerical methods are complex, the user-facing interface is designed to remain relatively simple. Default parameters are chosen to be physically sensible, while key routines remain accessible through a minimal number of function calls. The code also prioritizes robustness and clear identification of numerical pathologies, allowing problematic behavior to be distinguished from the expected evolution of the system. 

\end{itemize}

At its core, \EMRICode{}evolves an eight-component state vector consisting of the four spacetime coordinates and four components of the contravariant velocity. In the absence of radiation reaction, this state is evolved according to the Kerr geodesic equations. When radiation reaction is enabled, the geodesic evolution is periodically interrupted by an update to the orbital constants of motion, after which the corresponding four-velocity is reconstructed and the integration continues.

\section{Code Overview/Numerical Framework}
\label{Code Overview}
\subsection{Geometric Units and Orbital Constants}

Throughout this work we adopt geometric units with $G = c = 1$ and normalize the SMBH mass to $M = 1$. This choice renders the equations scale-independent and greatly simplifies the numerical evolution; Although \EMRICode{}performs all internal computations in this geometric framework, the code also accepts inputs in either SI or CGS units and automatically converts them as needed before and after integration, ensuring consistency and ease of use across different unit systems.

In this setting, the test-particle mass $\mu$ coincides with the standard mass ratio $\mu/M \ll 1$. Bound geodesic motion in Kerr spacetime is characterized by three conserved quantities: the specific energy $E$, the axial component of angular momentum $L_z$, and the Carter constant $C$, which is derived from a higher-order symmetry in Kerr spacetime related but generally not equivalent to non-axial angular momentum. Together with the dimensionless spin parameter $a$ $(0\le a\le 1)$, these integrals of motion fully determine the trajectory of a test particle in the absence of radiation \citep{Gravitation, Schmidt2002, Paganini_Ruba_Oancea_2016}.

\subsection{Effective Potentials and Geodesic Equations}

In Boyer-Lindquist coordinates, the Kerr metric admits separable equations of motion through the introduction of radial and polar effective potentials \citep{Sago_et_al_2006},
\begin{equation}
P(r) = E(r^2 + a^2) - aL_z \\
\label{EffPot_P}
\end{equation}
\begin{equation}
R(r) = P(r)^2 - \Delta\left[r^2 + (L_z - aE)^2 + C \right] \\
\label{EffPot_R}
\end{equation}
\begin{equation}
\Theta(\theta) = \left[C - a^2(1 - E^2)\cos^2\theta\right]\sin^2\theta - L_z^2\cos^2\theta
\label{EffPot_T}
\end{equation}
where $\Delta = r^2 - 2r + a^2$.

Expressed in terms of Mino time $\lambda$, the equations of motion are
\begin{gather}
    \label{EqMots_start}
    \frac{dt}{d\lambda} = -a(aE\sin^2\theta - L_z) + \frac{(r^2 + a^2)}{\Delta}P(r) \\
    \left(\frac{dr}{d\lambda}\right)^2 = R(r) \\
    \left(\frac{d\cos\theta}{d\lambda}\right)^2 = \Theta(\theta) \\
    \frac{d\phi}{d\lambda} = -\left(aE - \frac{L_z}{\sin^2\theta}\right) + \frac{a}{\Delta}P(r)
    \label{EqMots_end}
\end{gather}

However, these are primarily useful for the initialization of a new 8-element state $[x^\mu, u^\mu]$ from predefined parameters; for the geodesic evolution of an initial state through Kerr space, \EMRICode{}relies on the geodesic equation
\begin{gather}
\label{eq:geodesic_equations}
\frac{d^2(x^\mu)}{d\tau^2} = -\Gamma^\mu_{\alpha\beta}u^\alpha u^\beta \\
\frac{d(x^\mu)}{d\tau} = u^\mu\\
\frac{d(u^\mu)}{d\tau} = -\Gamma^\mu_{\alpha\beta}u^\alpha u^\beta
\end{gather}
where $\Gamma^\mu_{\alpha\beta}$ are Christoffel connection terms that describe the Kerr space, in order to actually step forward through integration.

\subsection{Initialization}
\label{Initialization}

Initialization of a run is computationally straightforward, but \EMRICode{}allows several distinct pathways depending on the needs of the user. The dimensionless spin $a$ and mass ratio $\mu/M$ are required for every calculation, while the initial orbital configuration can be specified through several alternative sets of parameters in whichever form is most convenient.

\begin{itemize}

    \item \textbf{Orbital Constants.}
    Users may initialize an orbit by providing the conserved quantities $E$, $L_z$, and $C$, together with an initial spacetime position. In this case, the corresponding four-velocity is obtained directly from the equations of motion Eqs.~\ref{EqMots_start} - \ref{EqMots_end}. If no position is supplied, or the location supplied is not consistent with the conserved quantities, the code defaults to an equatorial point located a distance $p$ (the semilatus rectum) from the central mass, where $p$ is derived from the turning points of the radial potential $R(r)$. Note that once radiation reaction is included, these values are naturally no longer conserved over the course of an inspiral; nonethless, we continue to refer to $E$, $L_z$, and $C$ collectively as the orbital constants for consistency.

    \item \textbf{Tetrad Velocities.}
    In some situations it is more natural to specify an initial 4-velocity rather than the constants of motion. \EMRICode{}therefore allows users to define initial velocities in the Zero Angular Momentum Observer (ZAMO) tetrad frame, whose spatial basis vectors align with the $\hat{r}$, $\hat{\theta}$, and $\hat{\phi}$ directions. The user provides a position $(r, \theta, \phi)$ and a tetrad-frame four-velocity 
    \begin{equation}
        \label{eq:tetrad_components}
        u^{\tilde{\mu}} = [\gamma,\, \gamma \vec{v}]
    \end{equation}
    or equivalently a speed $\beta = \left|\vec{v}\right|$ and orientation angles $(\eta, \xi)$, where $\eta$ is measured from the radial direction in the $r$--$\phi$ plane and $\xi$ from the $\theta$ direction. The speed must satisfy $\beta \leq 1$ and $\gamma$ is the usual Lorentz factor $\gamma = (1 - \beta^2)^{-1/2}$. The code converts the provided tetrad components to Boyer--Lindquist components via
    \begin{equation}
        \label{eq:tetrad_conversion}
        u^{\mu} = \mathscr{E}^{\mu}{}_{\tilde{\mu}}\, u^{\tilde{\mu}}
    \end{equation}
    where the transformation matrix is \citep{Schnittman_2013}
    \begin{equation}
        \label{eq:tet2kerr}
        \mathscr{E}^{\mu}{}_{\tilde{\mu}} =
        \begin{bmatrix}
            \frac{1}{\alpha} & 0 & 0 & 0 \\
            0 & \sqrt{\frac{\Delta}{\rho^{2}}} & 0 & 0 \\
            0 & 0 & \frac{1}{\rho} & 0 \\
            \frac{\omega}{\alpha} & 0 & 0 & \frac{1}{\varpi}
        \end{bmatrix}
    \end{equation}
with conveniently defined values
    \begin{gather}
        \rho^{2} \equiv r^{2} + a^{2}\cos^{2}\theta\\
        \Delta \equiv r^{2} - 2r + a^{2}, \\
        \alpha^{2} \equiv \frac{\rho^{2}\Delta}{\rho^{2}\Delta + 2r(r^{2} + a^{2})}\\
        \omega \equiv \frac{2ra}{\rho^{2}\Delta + 2r(r^{2} + a^{2})}, \\
        \varpi^{2} \equiv \frac{\rho^{2}\Delta + 2r(r^{2} + a^{2})}{\rho^{2}} \sin^{2}\theta.
    \end{gather}

    \item \textbf{Observable (Kepler-like) Parameters.}
    Alternatively, users may initialize an orbit using parameters analogous to those in classical celestial mechanics: the semimajor axis $r_{0}$, eccentricity $e$, and inclination $\theta_{-}$, with $\theta_{-} = \pi/2$ corresponding to an equatorial orbit, $\theta_{-} = 0$ to a polar orbit, and negative values of $\theta_{-}$ indicating retrograde motion. The corresponding constants of motion are then obtained using the mapping of \citet{Schmidt2002}.

    \item \textbf{Predefined Trajectory.}
    Finally, \EMRICode{}can launch directly from a fully specified state vector (four space-time coordinates and four velocity components in Boyer-Lindquist). This option is particularly useful when continuing, refining, or comparing to a pre-existing simulation, as it bypasses all conversion steps.

\end{itemize}

\subsection{Adaptive Geodesic Integration}

We integrate and evolve the geodesic equations using a fifth-order Cash–Karp Runge–Kutta integrator with embedded fourth-order error estimates, enabling dynamic adaptive timestep behavior based on local truncation error \citep{Schnittman_2004,Schnittman_2013}. 

The local truncation error is estimated using the embedded fourth- and fifth-order Runge-Kutta solutions $s_4$ and $s_5$, respectively. The relative timestep error is then computed as 
 
\begin{gather}
    \Delta s = s_4 - s_5, \\
    \epsilon = \sqrt{\frac{1}{8}(\Delta s \cdot \Delta s)}
\end{gather}

The timestep is then updated via
\begin{gather}
    \Delta\tau_1 = \min\!\left(0.04\pi (x^r)^{1.5},\;
    \Delta\tau_0 \left|\frac{\epsilon_0}{\epsilon + 0.01\epsilon_0}\right|^{0.2}\right),
\end{gather}
where $\Delta\tau_0$ and $\epsilon_0$ indicate the original timestep and the user-specified geodesic error tolerance, respectively. This particular formulation is used to ensure that every orbit consists of at least $\sim 50$ time steps, which we find is sufficient for the numerical derivatives used in the multipole calculation of the next Section.

To enforce the normalization condition $g_{\mu\nu} u^\mu u^\nu = -1$, we occasionally find it necessary to rescale $u^t$ by solving
\begin{gather}
g_{tt}(\kappa u^t)^2 + 2g_{t\phi}u^\phi(\kappa u^t)+g_{ij}u^i u^j = -1,
\end{gather}
yielding
\begin{gather}
\kappa = \frac{-2g_{t\phi}u^\phi u^t + \sqrt{(2g_{t\phi}u^\phi u^t)^2 - 4g_{tt}(u^t)^2(g_{ij}u^iu^j + 1)}}{2g_{tt}(u^t)^2}.
\end{gather} 

In practice, this renormalization procedure is required only infrequently, occurring primarily for highly eccentric trajectories or when exceptionally stringent error tolerances are imposed. Even in the most extreme cases considered here, fewer than approximately 1 in 20 integration steps require renormalization, while for typical trajectories the fraction is substantially smaller. Furthermore, the correction itself is always small, with the rescaling factor satisfying $|1-\kappa| < \epsilon_0$.

A further numerical complication arises for trajectories passing near the poles of the Boyer-Lindquist coordinate system. Although these coordinate singularities do not correspond to physical singularities, the adaptive timestep algorithm tends to select progressively smaller timesteps as the trajectory approaches $\theta=0$ or $\theta=\pi$, leading to inefficient integration and, in some cases, numerical stagnation.

To mitigate this behavior, we employ a specialized pole-crossing procedure for near-polar trajectories. When a particle approaches a pole sufficiently closely and the adaptive timestep exhibits sustained collapse, we assume that the remaining angular separation is small enough that the trajectory may be approximated locally by a straight-line crossing through the polar region. The corresponding azimuthal displacement is then determined geometrically by treating the motion as a chord on the sphere defined by the current position and the projected closest approach to the pole (the orbital inclination, determined by the effective potential in Eq.~\ref{EffPot_T}), allowing the trajectory to be mapped to the opposite side of the pole while preserving the direction of motion. The elapsed coordinate time is estimated from the displacement and local angular velocity, and the sign of $u^\theta$ is reversed to reflect the completed crossing. Because this procedure is invoked only when the trajectory is already extremely close to a coordinate singularity, it introduces negligible physical error while substantially improving numerical efficiency and stability for highly inclined or nearly polar orbits. 

\subsection{Radiation Reaction}

Radiation reaction is implemented as a separate correction to the geodesic evolution, allowing different prescriptions for the secular evolution of the orbital constants to be substituted without modifying the underlying geodesic integrator. To model radiation-reaction-driven inspiral, we update the orbital constants once per radial orbit using the leading-order quadrupole formalism of \citet{Peters1964}. The time-averaged energy and angular momentum fluxes due to radiation reaction are
\begin{gather}
    \label{quadrupole_def}
    Q_{ij} = \mu\left(x_i x_j - \frac{1}{3}r^2\delta_{ij}\right), \\
    \frac{dE}{dt} = -\frac{1}{5}\left[\left(\dddot{Q}_{ij}\right)^2 - \frac{1}{3}(\dddot{Q}_{ii}\dddot{Q}_{jj})\right], \\
    \frac{dL_i}{dt} = -\frac{2}{5}\epsilon_{ijk} (\ddot{Q}_{jm}\dddot{Q}_{km}).
\end{gather}
These equations are integrated over a single radial geodesic period, giving the small changes to the orbital parameters $\delta E$, $\delta L_i$.

Because the adaptive timestep scheme produces nonuniform sampling in the affine parameter, the resulting trajectories are subsequently tabulated and resampled using cubic splines. This approach preserves the accuracy benefits of adaptive integration while enabling efficient post-processing and evaluation of orbital quantities at arbitrary points in the orbit.

These fluxes provide direct evolution equations for the orbital energy $E$ and the Cartesian angular momentum vector $L_i$. Determining the evolution of the Carter constant $C$ is less straightforward, as no corresponding gauge-invariant flux exists in the general Kerr case. Consequently, several approximate prescriptions for $\delta C$ were implemented and compared. Because these prescriptions are formulated in terms of the components of $L_i$, they require transforming between Boyer-Lindquist coordinates and a Cartesian-like coordinate system adapted from the asymptotically flat limit of the Kerr geometry. Although this transformation reduces to the familiar spherical-to-Cartesian mapping in the Schwarzschild limit, the general Kerr metric does not implicitly conserve the non-axial components $L_x$ and $L_y$ \citep{Gravitation, Paganini_Ruba_Oancea_2016}.  Furthermore, while the fluxes provide four independent evolution equations, corresponding to $E$ and the three components of $L_i$, bound Kerr geodesics are characterized by only three independent constants of motion. Recovering the perturbed orbital state therefore constitutes an overdetermined problem, necessitating the use of approximate reconstruction procedures whose accuracy and numerical stability are investigated below. The performance and consistency of these prescriptions are evaluated in Section~\ref{sec:DaV}.

\subsubsection{Inclination-Preserving Approximation}
\label{inc0}

A common approximation assumes that the orbital inclination remains nearly constant throughout the inspiral. This behavior has been found in previous studies of extreme mass-ratio inspirals, where inclination evolution is generally much slower than the evolution of the remaining orbital parameters. Establishing a commonly used definition of inclination:

\begin{gather}
    \cos\iota = \frac{L_z}{\sqrt{L_z^2 + C}} \approx \sin{\theta_-},
\end{gather}

and imposing $\dot{\iota}=0$ yields

\begin{gather}
    \delta C = 2C\frac{\delta L_z}{L_z} ~\label{eq:[con-inc-method-equation]}
\end{gather}

This expression becomes singular for exactly polar orbits ($L_z=0$). In these cases, the formulation of \citet{Glampedakis2002} is used to express the evolution in terms of the remaining angular momentum components and the energy flux. We combine the definitions of total square angular momentum from Schwarzschild spacetime
\begin{gather}
    L^2 = p_{\theta}^2 + p_{\phi}^2(\sin{\theta})^{-2}
\end{gather}
and the Carter constant as expressed in \citet{Gravitation}
\begin{gather}
    C = p_{\theta}^2 + L_z^2\cot^2{\theta} + a^2(1-E^2)\cos^2{\theta}
\end{gather}
to obtain the approximate expression:
\begin{align}
    C &= L^2-L_z^2+a^2(1 - E^2)\cos^2{\theta}  \notag\\
    &= L_x^2 + L_y^2+a^2(1 - E^2)\cos^2{\theta} \, .
\end{align}
Because this definition of $C$ remains approximately constant over the course of geodesic orbit, we may choose to evaluate it in the equatorial plane at $\theta=\frac{\pi}{2}$, eliminating the cosine term; we then differentiate to obtain our final expression
\begin{equation}
    \delta C = 2L_x\delta L_x + 2L_y\delta L_y.
    ~\label{eq:[con-inc-method-equation-pole]}
\end{equation}

Because the Cartesian components of the angular momentum are not conserved quantities in Kerr spacetime and do not admit simple closed-form expressions in Boyer-Lindquist coordinates, they are reconstructed numerically from the instantaneous trajectory. The particle position and four-velocity are first transformed to a Cartesian-like coordinate system, after which the orbital angular momentum direction is obtained from the cross product of the position and velocity vectors. This direction is subsequently rescaled so that its magnitude satisfies $|\mathbf{L}|^2=L^2$, yielding estimates of $L_x$, $L_y$, and $L_z$ consistent with the total angular momentum. Because this procedure relies on a coordinate transformation rather than an exact Kerr symmetry, it introduces a small reconstruction error, particularly for strong-field, high-spin, and highly inclined orbits. The magnitude and consequences of this approximation are examined in Section~\ref{sec:DaV}.

\subsubsection{Potential-Based Approximation}

An alternative prescription is obtained from the radial effective potential $R(r)$. Evaluating the radial potential at a turning point and solving for the Carter constant gives

\begin{gather}
C = \frac{P(r)^2}{\Delta} - r^2 - (aE-L_z)^2.
\end{gather}

Taking the total differential yields

\begin{gather}
\delta C =
\frac{\partial C}{\partial E}\delta E
+
\frac{\partial C}{\partial L_z}\delta L_z
+
\frac{\partial C}{\partial r}\delta r,
\end{gather}

where

\begin{gather}
\frac{\partial C}{\partial E} 
= 2\frac{(r^2 + a^2)P(r)}{\Delta} - 2a(aE-L_z) \\
\frac{\partial C}{\partial L_z} 
= -2a\frac{P(r)}{\Delta} + 2(aE-L_z) \\
\frac{\partial C}{\partial r} 
= \frac{2P(r)}{\Delta^2}(2Er\Delta + (1-r)P(r)) - 2r \\
\end{gather}

Because the locations of the radial turning points are themselves determined by the evolving orbital constants, $\delta r$ is not an independent perturbation. A fully self-consistent treatment would require simultaneously solving for the evolution of both the turning points and the Carter constant, resulting in a coupled perturbation problem. In the present work, we instead adopt a first-order approximation in which the explicit radial contribution is treated as a higher-order correction. Accordingly, $\delta C$ is estimated by retaining only the terms proportional to the directly computed fluxes and integrating them along the orbit,

\begin{gather}
\delta C = \int_{t_0}^{t_f}
\left( \frac{\partial C}{\partial E}\frac{dE}{dt}
+\frac{\partial C}{\partial L_z}\frac{dL_z}{dt}\right)\bigg|_{r=r(t)}
dt. ~\label{eq:[radial-method-equation]}
\end{gather}

This approximation provides a direct relationship between the quadrupole fluxes and the evolution of the Carter constant while avoiding an explicit determination of the turning-point motion.

\subsubsection{Tetrad Least-Squares Approximation}

A third approach estimates the evolution of the Carter constant by directly reconstructing a perturbed orbital state consistent with the computed energy and angular momentum losses.

The particle four-velocity is first transformed into the locally non-rotating (ZAMO) tetrad frame, where it takes the form

\begin{gather}
u^{\tilde{\mu}} =
[\gamma,\gamma v^{\tilde{i}}].
\end{gather}

Working in the tetrad frame guarantees that any normalized four-velocity corresponds to a physically admissible timelike trajectory, allowing the orbital constants to be adjusted through perturbations of only the three independent spatial velocity components.

Assuming these perturbations are small, the corresponding changes in the orbital constants are approximated by the linear system

\begin{gather}
    \begin{bmatrix}
        \partial E/\partial   v^{\tilde{r}} & \partial E/\partial   v^{\tilde{\theta}} & \partial E/\partial   v^{\tilde{\phi}} \\
        \partial L_x/\partial v^{\tilde{r}} & \partial L_x/\partial v^{\tilde{\theta}} & \partial L_x/\partial v^{\tilde{\phi}} \\
        \partial L_y/\partial v^{\tilde{r}} & \partial L_y/\partial v^{\tilde{\theta}} & \partial L_y/\partial v^{\tilde{\phi}} \\
        \partial L_z/\partial v^{\tilde{r}} & \partial L_z/\partial v^{\tilde{\theta}} & \partial L_z/\partial v^{\tilde{\phi}} 
    \end{bmatrix}
    \begin{bmatrix}
        \delta v^{\tilde{r}}\\
        \delta v^{\tilde{\theta}}\\
        \delta v^{\tilde{\phi}}
    \end{bmatrix} = 
    \begin{bmatrix}
        \delta E\\
        \delta L_x\\
        \delta L_y\\
        \delta L_z
    \end{bmatrix}
    \label{eq:[lsq-method-equation]}
\end{gather}

where we label the $4\times3$ Jacobian matrix $A$. Given the coordinate-dependent construction of $L_x$ and $L_y$ and the multiple transformations relating these quantities to the tetrad-frame velocity, the velocity derivatives in $A$ were evaluated using central finite differences rather than derived analytically. Each column is obtained by perturbing a single tetrad velocity component while holding the remaining components fixed, transforming the perturbed velocity back to Boyer-Lindquist coordinates, and recomputing the corresponding energy and angular momentum components with the same method described in Section \ref{inc0}. The rows of $A$ therefore represent the local sensitivities of $[E,L_x,L_y,L_z]$ to perturbations in each tetrad velocity component.

Because four target quantities are specified but only three independent velocity perturbations are available, the system is overdetermined. Rather than enforcing an exact solution, we determine the velocity perturbation that minimizes the residual in a least-squares sense using the Moore--Penrose pseudoinverse \citep{Moore_1920, Penrose_1955},

\begin{gather}
\mathbf{x} = (A^{T}A)^{-1}A^{T}\mathbf{b},
\end{gather}

where $\mathbf{x}$ contains the tetrad velocity perturbations and $\mathbf{b}$ contains the desired changes in energy and angular momentum. The perturbed velocity is then transformed back to Boyer-Lindquist coordinates, from which the updated orbital constants, including the Carter constant, are computed directly. Additional implementation details, including the numerical construction of the Jacobian and conditioning of the least-squares system, are discussed in Appendix~A.

\subsection{Exit Conditions and Output Structure}

By default, the simulation terminates either upon detecting a plunge into the central black hole or after reaching a maximum of $10^{7}$ integration steps, whichever occurs first. Users may optionally specify alternative termination criteria, including thresholds on coordinate time, minimum or maximum radial position, the number of completed radial or azimuthal oscillations, or specified Keplerian orbital elements. The simulation may also be stopped manually at any point; in all cases, the code returns the accumulated data up to the moment of termination.

The final output is returned as a structured dictionary containing both the original input parameters and the raw trajectory data. In addition to the full spacetime trajectory, the output includes the time series of conserved quantities, instantaneous Keplerian elements, true anomaly, Mino time, and the locations of ascending and descending nodes, as well as the azimuthal angle of periapsis.

Because many output quantities are derived from a common set of underlying variables, the framework also provides optional compression and reconstruction utilities for long-term data storage. These routines retain only the information necessary to reproduce the full set of derived quantities when the simulation output is reloaded.

\section{Demonstration and Validation}
\label{sec:DaV}

The parameter space of Extreme Mass-Ratio Inspirals is extensive, and a comprehensive exploration of the full range of orbital configurations is beyond the scope of this paper, which is intended to demonstrate the capabilities and behavior of \EMRICode{.}Instead, we select a set of representative trajectories designed to sample qualitatively distinct regions of the EMRI parameter space and illustrate the accessible range of orbital behaviors. These trajectories serve as recurring case studies throughout this section, while broader parameter sweeps are used where systematic characterization of the code's behavior is more informative than individual examples.

We consider four representative orbital configurations:
\begin{enumerate}
    \item \textbf{Quasi-Circular Inclined Orbit.}
    This case provides a baseline test of long-term phase coherence and numerical stability in the absence of significant radial structure. The motion is dominated by smooth azimuthal and polar oscillations, making it particularly sensitive to cumulative integration error. The orbit is initialized with a semilatus rectum $p=10$, eccentricity $e=10^{-3}$, maximum inclination $\theta_{-}=-60^\circ$, and dimensionless spin $a=0.9$. The negative value of $\theta_{-}$ corresponds to a retrograde orbit relative to the black-hole spin.
    
    \item \textbf{Equatorial Zoom--Whirl Orbit.}  
    'Zoom--whirl' orbits are characterized by extended azimuthal motion near periapsis, followed by excursions to much larger radii; more analytically, they can be characterized by their proximity to the separatrix between bound and plunging motion. This configuration probes the strong-field relativistic regime and places stringent demands on adaptive timestep control due to both the rapid variation in local dynamical timescales and the proximity of the periapse to the black hole horizon. The orbit is initialized with $p=2.600$, $e=0.95$, $\theta_{-}=90^\circ$, and $a=0.98$, corresponding to an equatorial orbit. In this configuration, the secondary body completes 6 azimuthal orbits in the time it takes to finish one radial cycle.
    
    \item \textbf{Generic Eccentric and Inclined Orbit.}  
    This case combines nonzero eccentricity and inclination and therefore exercises the full three-frequency structure of generic Kerr geodesic motion. It samples a representative region of the EMRI parameter space and provides a comprehensive test of the integrator under simultaneously nontrivial radial, polar, and azimuthal motion. The orbit is initialized with $p=12$, $e=0.35$, $\theta_{-}=-30^\circ$, and $a=0.8$, corresponding to a moderately eccentric, retrograde orbit.
    
    \item \textbf{Near-Polar Orbit.}  
    Orbits with inclinations approaching the polar configuration probe regions where the polar motion brings the trajectory close to the coordinate poles, potentially introducing numerical sensitivities associated with the angular coordinates. This configuration therefore provides a test of the treatment of the Carter constant and the robustness of the integrator near the coordinate poles. The orbit is initialized with $p=9.831$, $e=0.4211$, $\theta_{-}=5.113^\circ$, and $a=0.95$.
    
\end{enumerate}

Together, these examples span a wide range of orbital behavior while remaining computationally tractable and physically relevant. The representative trajectories are complemented by systematic parameter studies throughout this section, allowing the behavior of \EMRICode{}to be assessed both for individual orbital configurations and across broader regions of parameter space.

\subsection{Geodesic Validation}
\label{sec:geo_val}

All trajectories shown in this section are evolved as pure geodesics, with radiation reaction disabled, in order to isolate the performance of the geodesic integrator and to assess long-term conservation of the orbital constants. Each trajectory is integrated for a duration corresponding to roughly 500 radial orbits. Short segments of each trajectory are shown for visualization purposes, while the full integrations are used for spectral analysis and conservation diagnostics. The most stringent trajectories, considered the most 'accurate' for the purposes of comparison, are evaluated with a tolerance of $\epsilon_0=10^{-15}$.

\subsubsection{Orbital Morphology}

\begin{figure}
    \centering
    \includegraphics[scale=0.57]{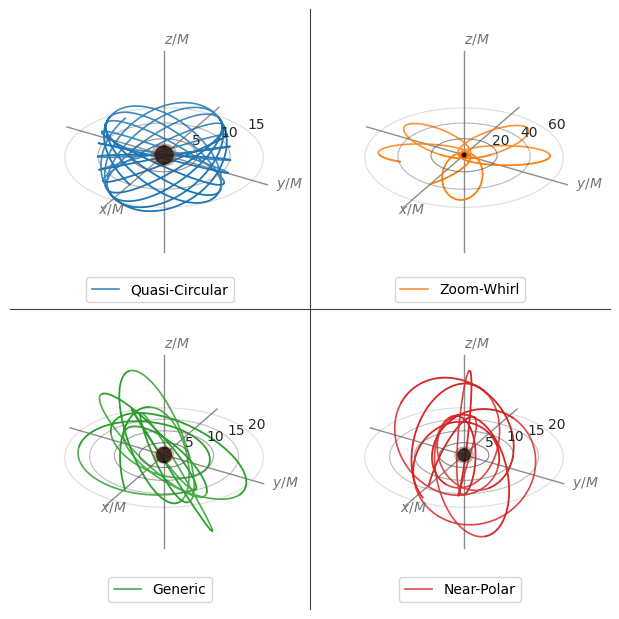}
    \caption{The first 5 radial orbits of each of the representative trajectories.}
    \label{fig:trajectories}
\end{figure}

Figure~\ref{fig:trajectories} shows short segments of each trajectory plotted in three dimensions. For clarity, only five radial orbital cycles are displayed in each panel, highlighting the geometric structure of the motion without obscuring individual features. The quasi-circular orbit exhibits smooth precession, while the Zoom--Whirl trajectory displays tightly wound azimuthal motion near periapsis. The generic orbit traces a complex, non-repeating path characteristic of triperiodic Kerr motion, and the polar orbit shows large-amplitude excursions in the polar direction.

Although these visualizations represent only a small fraction of the full integration time, they provide intuition for the qualitative differences between orbital families that later manifest in the spectral and conservation analyses.

\subsubsection{Orbital Constant Conservation}
\label{sec:orb_con_con}

To assess the numerical conservation of the geodesic integrator, we monitor the fractional error in the orbital constants. For equatorial orbits, such as our Zoom--Whirl example, $C$ vanishes and makes obtaining the fractional error impossible; in lieu of $C$, we track the Carter-related quantity
\begin{gather}
    Q = C + (aE-L_z)^2
\end{gather}
along with $E$ and $L_z$ throughout each integration. For a pure geodesic, these quantities are conserved, so any observed evolution provides a direct measure of numerical error
\begin{gather}
    \frac{\Delta X}{X_0} = \frac{X(t)-X_0}{X_0},
\end{gather}
where $X$ denotes each of the orbital constants.

\begin{figure}
    \centering
    \includegraphics[scale=0.62]{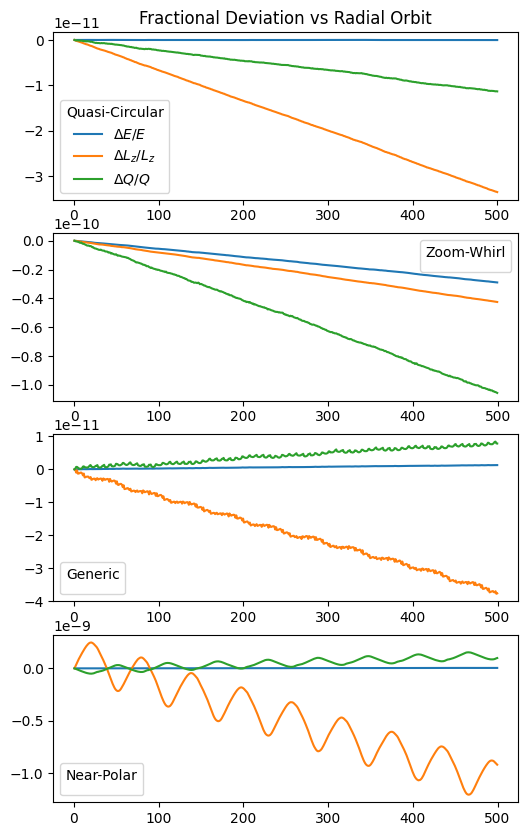}
    \caption{Fractional deviations in the orbital energy $E$, axial angular momentum $L_z$, and Carter-related quantity $Q$ as a function of the number of radial orbits for the four representative geodesic trajectories. }
    \label{fig:frac_dev}
\end{figure}

Figure~\ref{fig:frac_dev} shows the fractional deviations in $E$, $L_z$, and $Q$ for the four representative trajectories over the full integrations, expressed in terms of the number of completed radial orbits. The integrations are all performed using the default geodesic integration tolerance of $10^{-15}$.

All four trajectories remain well-conserved over the full integration. The maximum fractional deviations are summarized in Table~\ref{tab:geodesic_conservation}. The Zoom--Whirl trajectory exhibits the largest deviations in both energy and angular momentum, with maximum fractional errors of approximately $2.9\times10^{-11}$ and $4.2\times10^{-11}$, respectively. The Zoom--Whirl and Near-Polar trajectories have similar values for their maximum deviations in $Q$, but the Near-Polar deviation is slightly larger at roughly $1.5\times10^{-10}$. Across any given trajectory, the deviation in energy tends to be the lowest.

\begin{table}
\centering
\caption{Maximum Fractional Deviations in Orbital Constants}
\begin{tabular}{lccc}
\hline
Orbit & $\max|\Delta E/E|$ & $\max|\Delta L_z/L_z|$
     & $\max|\Delta Q/Q|$ \\
\hline
Quasi-circular & $3.546\times10^{-14}$ & $3.349\times10^{-11}$ & $1.132\times10^{-11}$ \\
Zoom--Whirl    & $2.983\times10^{-11}$ & $4.249\times10^{-11}$ & $1.056\times10^{-10}$ \\
Generic        & $1.319\times10^{-12}$ & $3.777\times10^{-11}$ & $8.412\times10^{-12}$ \\
Near-Polar   & $4.239\times10^{-12}$ & $1.208\times10^{-9}$ & $1.533\times10^{-10}$ \\
\hline
\end{tabular}
\label{tab:geodesic_conservation}
\end{table}

A common feature of the conservation diagnostics is that the fractional deviation in $L_z$ tends to grow more rapidly than those of $E$ and $Q$, particularly for the quasi-circular, generic, and Near-Polar trajectories. This behavior is partly a consequence of the relative magnitudes of the quantities being monitored: a comparable absolute numerical error produces a larger fractional deviation when the corresponding conserved quantity is smaller. This effect is especially important for the Near-Polar trajectory, for which $L_z$ is small compared with the total angular momentum.

The Near-Polar trajectory also exhibits the clearest temporal structure in its conservation errors. In addition to a slowly varying secular component, the deviations contain a bounded oscillatory component that is apparent in all three orbital constants.  The Generic trajectory shows weaker indications of similar structure, although its oscillatory component is less clearly distinguished from numerical fluctuations. The Quasi-Circular trajectory exhibits comparatively little obvious long-period structure, while the Zoom--Whirl trajectory is dominated by relatively small fluctuations over the limited number of radial cycles covered by the integration. The absence of an obvious long-period component in the latter case may therefore reflect the shorter effective sampling of the orbital evolution rather than a fundamentally different numerical behavior.

The larger errors observed for the Near-Polar trajectory may be consistent with the increased numerical difficulty associated with its orbital geometry. Trajectories approaching the poles of Boyer--Lindquist coordinates require the integrator to resolve rapid changes in the coordinate representation of the motion while maintaining the normalization and conserved quantities. The larger fractional deviation in $L_z$ is further amplified by the relatively small magnitude of $L_z$ for this orbit. However, an investigation of the trajectory reveals that it lies near a resonance between the radial and polar frequencies such that $\Omega_r/\Omega_{\theta} \approx 2/3$, and that the observed frequency of the oscillation in error is $f_{osc} \approx |2\Omega_{\theta} - 3\Omega_r|$. Limited exploration of other trajectories around this and other low-order resonances seem to show similar behavior, indicating some unexpected coupling between the numerical integrator and resonant orbits. Nevertheless, the accumulated deviations remain small even in this most demanding representative case.

\begin{figure}
    \centering
    \includegraphics[scale=0.6]{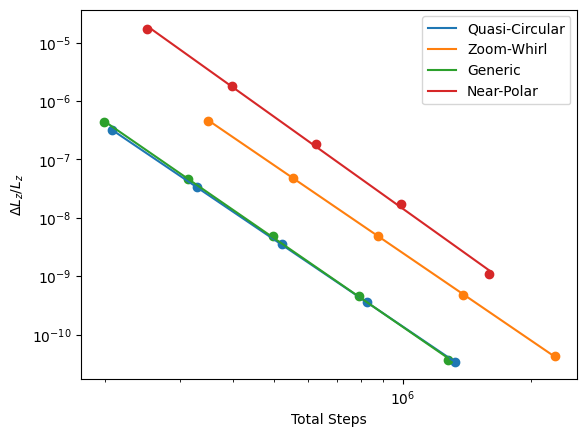}
    \caption{Final fractional deviation in $L_z$ as a function of the total number of integration steps for the representative geodesic trajectories at different integration tolerances. The observed scaling ($\epsilon \sim N^{-4}$) is consistent with the expected convergence of the underlying Cash-Karp integration scheme.}
    \label{fig:tolerance_convergence}
\end{figure}

To examine the convergence of the numerical conservation with integration tolerance, the representative trajectories were also evolved using tolerances of $10^{-14}$, $10^{-13}$, $10^{-12}$, and $10^{-11}$. The accumulated conservation errors increase systematically as the tolerance is relaxed. Figure~\ref{fig:tolerance_convergence} shows the resulting dependence of the final conservation error on the number of integration steps.

For each representative trajectory, relaxing the integration tolerance results in fewer integration steps and a corresponding increase in the accumulated conservation error. The approximately parallel trends between the different trajectories indicate that the dominant secular component of the numerical error follows a similar convergence behavior across the tested orbital configurations. The observed scaling is consistent with the expected convergence of the fifth-order Cash--Karp integration scheme, while the differing vertical offsets reflect the different numerical demands of the individual trajectories.

The detailed behavior of the Near-Polar trajectory provides an additional indication that the observed deviations are predominantly numerical. Its oscillatory component remains at approximately the same characteristic frequency when the integration tolerance is varied, and while its amplitude does increase with looser tolerances, the magnitude of the secular component grows rapidly enough to quickly dominate the overall error. We consequently do not interpret this structure as physical evolution of the orbital constants. Instead, its  dependence on the integration tolerance are consistent with a component of the numerical error.

Overall, these geodesic tests demonstrate that \EMRICode{}maintains the conserved orbital quantities to high accuracy over integration times comparable to relevant observational timescales. Even for the Near-Polar trajectory, which represents the most numerically demanding of the selected cases, the maximum fractional deviations remain below approximately $10^{-9}$ at the default integration tolerance. The systematic increase in error with relaxed tolerance, together with the small accumulated deviations at the default tolerance, indicates that the geodesic integrator provides adequate numerical accuracy for the subsequent radiation-reaction calculations.

These tests characterize the accuracy of the underlying geodesic integration in the absence of radiation reaction. When radiation reaction is enabled, the orbital constants are no longer conserved; instead, their secular evolution is determined by the prescribed energy and angular-momentum fluxes. Consequently, the conservation diagnostics used here are not directly applicable to the inspiral calculations. Validation of the radiation-reaction implementation instead requires assessing whether the prescribed fluxes produce physically consistent changes in the orbital parameters and whether those changes remain numerically stable over long integrations. Section~\ref{sec:rrs} therefore examines the prescriptions used to determine the evolution of the Carter constant and their effect on the resulting inspiral trajectories.

\subsubsection{Spectral Content}
\label{sec:spectral}

Although the geodesic trajectories considered above do not include radiation reaction and therefore do not represent self-consistent inspirals, their orbital motion provides a useful diagnostic of the frequency content that would enter an associated gravitational-wave signal. To examine this structure, for each trajectory we compute the Fourier transform of the quadrupole waveform polarizations, which are calculated like so in accordance with \citet{Gravitation}: 
\begin{gather}
    P = I_3 - \hat{n} \otimes \hat{n} \\
    h_{TT} = \frac{2}{d}\big(P\ddot{Q}P - \frac{1}{2}\Tr[P\ddot{Q}\big]P\big) \\
    h_{+} = \frac{1}{2}\big(\hat{e_x} \otimes \hat{e_x} - \hat{e_y} \otimes \hat{e_y} \big)h_{TT} \\
    h_{\times} = \big(\hat{e_x} \otimes \hat{e_y} \big) h_{TT}
\end{gather}
where \{$\hat{e_x}, \hat{e_y}, \hat{n}$\} form an orthonormal basis with $\vec{d} = d \hat{n}$ as the distance vector from the source to the observer; for each trajectory, the distance vector is set such that the observer is an arbitrary distance of $10^5$ gravitational radii away in the direction of the central body's angular momentum vector. This analysis is not intended to predict the detailed gravitational-wave spectrum of an EMRI, but rather to determine whether the numerical trajectories reproduce the characteristic frequency structure expected for bound Kerr orbits.

Bound Kerr geodesics possess three fundamental orbital angular frequencies, $\Omega_r$, $\Omega_\theta$, and $\Omega_\phi$, associated with radial, polar, and azimuthal motion, respectively. The frequency-domain structure of an EMRI waveform can therefore be decomposed into harmonics with frequencies of the form
\begin{gather}
    \omega_{mkn} = m\Omega_\phi + k\Omega_\theta + n\Omega_r,
\end{gather}
where $m$, $k$, and $n$ are integers \citep{Katz2021}. The resulting spectrum is consequently expected to consist of discrete combinations of the three fundamental orbital frequencies, with the relative strength of individual harmonics depending on the orbital configuration and observer orientation. The mode numbers are not restricted to positive values, as the Fourier decomposition permits each fundamental frequency to contribute with either sign; equivalently, the resulting harmonics can be viewed as sidebands produced by the modulation and mixing of the fundamental frequencies. This makes the spectral content a useful qualitative diagnostic of the numerical orbital dynamics. In this section, the angular frequencies $\Omega_i$ are replaced with the cyclic frequencies $f_i = \frac{1}{2\pi}\Omega_i$ for ease of analysis; see Table~\ref{tab:char_freq_inits} for an accounting of observed frequencies.

\begin{table}
\centering
\caption{Characteristic frequencies}
\begin{tabular}{lccc}
\hline
Orbit & $f_{\phi}$ & $f_{\theta}$ & $f_r$\\
\hline
Quasi-Circular & $5.104\times 10^{-3}$ & $5.442\times 10^{-3}$ & $2.255\times 10^{-3}$\\
Zoom--Whirl & ${6.124\times 10^{-3}}^*$ & $2.655\times 10^{-3}$ & $9.609\times 10^{-4}$\\
Generic & $3.266\times 10^{-3}$ & $3.408\times 10^{-3}$ & $2.179\times 10^{-3}$\\
Near-Polar & $4.341\times 10^{-3}$ & $4.084\times 10^{-3}$ & $2.707\times 10^{-4}$\\
\hline
\multicolumn{4}{l}{\scriptsize $^*$The periapse-associated azimuthal frequency $f_{\phi p} = 6.350\times 10^{-2}$}
\end{tabular}
\label{tab:char_freq_inits}
\end{table}

For this analysis, the orbital coordinates are interpolated onto a uniformly sampled coordinate-time grid before constructing the quadrupole moment and computing its second time derivative. The resulting transverse-traceless quadrupole waveform is then Fourier-transformed. The same procedure is applied to each of the four representative trajectories described above, allowing their spectral structures to be compared directly.

\begin{figure}
    \centering
    \includegraphics[scale=0.49]{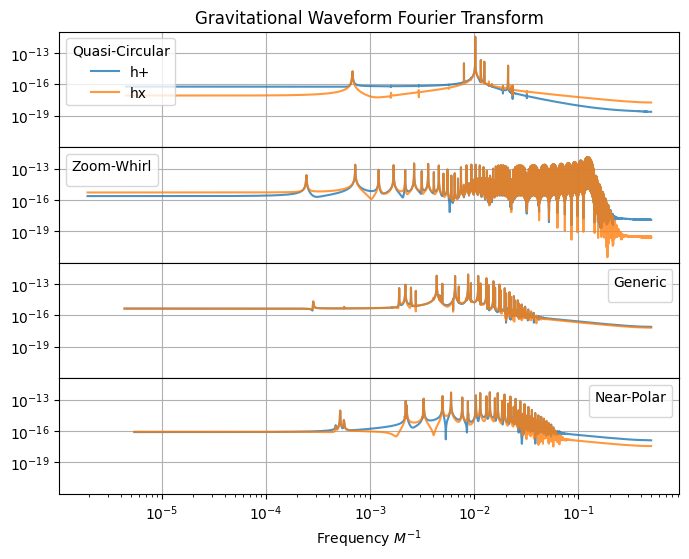}
    \caption{Fourier amplitudes of the quadrupole-based proxy waveforms for
    the four representative geodesic trajectories. Frequencies are given in
    geometric units of $M^{-1}$.}
    \label{fig:geodesic_spectra}
\end{figure}

Figure~\ref{fig:geodesic_spectra} shows the Fourier amplitudes of the four representative trajectories over the full sampled frequency range. The spectra exhibit substantial differences between the orbital configurations, while retaining the discrete and clustered structure expected from multi-frequency Kerr motion. The quasi-circular trajectory is dominated by a single narrow feature at $1.02\times10^{-2} M^{-1}$, with substantially less power in its subdominant harmonics. This frequency corresponds closely to $2f_\phi$, consistent with the expected dominance of the azimuthal harmonic at twice the orbital frequency for circular Kerr orbits \citep{Hughes2001}. In contrast, the eccentric and inclined trajectories distribute their power among a much larger number of harmonics.

The Zoom--Whirl trajectory exhibits the broadest high-power frequency structure of the four cases. Several groups of closely spaced peaks extend to frequencies of approximately $4\times10^{-2}/M$, reflecting the rapid azimuthal motion associated with the repeated periapsis passages. Above this region, the spectrum falls sharply and the two polarizations become increasingly sensitive to numerical noise. The generic and near-polar trajectories likewise exhibit clustered harmonic structure, but their power is concentrated more strongly toward the portion of the spectrum corresponding to the fundamental orbital frequency.

\begin{figure}
    \centering
    \includegraphics[scale=0.49]{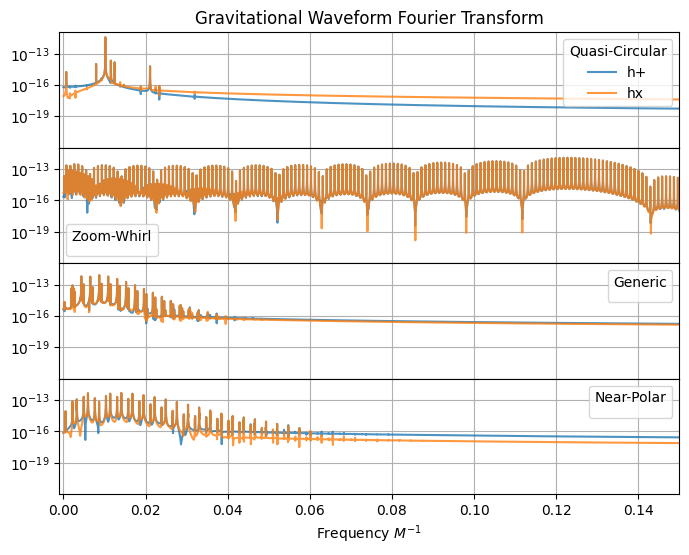}
    \caption{Fourier amplitudes of the representative trajectories on a
    linear frequency scale, restricted to $0 \leq f \leq 0.15 M^{-1}$. The
    linear scale emphasizes the clustering and spacing of the dominant
    spectral features.}
    \label{fig:geodesic_spectra_linear}
\end{figure}

\begin{table}
\centering
\caption{Fourier Spectral Binning Power Distribution}
\begin{tabular}{lccc}
\hline
Orbit & 50\% & 90\% & 99\%  \\
\hline
Quasi-circular & 1 & 1 & 1 \\
Zoom--whirl    & 12 & 110 & 1106 \\
Generic        & 4 & 23 & 206 \\
Near-polar     & 6 & 47 & 459\\
\hline
\end{tabular}
\label{tab:geodesic_spectral_binning}
\end{table}

The concentration of spectral power can be quantified by counting the number of frequency bins required to contain specified fractions of the total power. The results are shown in Table~\ref{tab:geodesic_spectral_binning}. The Quasi-Circular trajectory is particularly concentrated, with over 90\% of its total power contained within 1 bin. The remaining trajectories exhibit substantially broader spectral distributions. The Generic trajectory retains a relatively concentrated dominant spectrum, requiring only four bins to contain 50\% of the total power and 23 bins to contain 90\%, but requires 206 bins to capture 99\% of the power. The Zoom--Whirl trajectory exhibits the broadest spectral distribution, with just over 1100 bins required to contain 99\% of the power.

These results quantify the concentration of spectral power but do not, by themselves, characterize the structure of the spectrum. In particular, the Generic and Near-Polar trajectories exhibit similar spectral shapes upon visual inspection, with the Near-Polar trajectory displaying a slightly broader distribution at a different peak. Although the Zoom--Whirl trajectory requires substantially more frequency bins to capture the same fractions of total power, this difference cannot be interpreted simply as a further broadening of an otherwise similar spectrum. Its spectral distribution exhibits qualitatively distinct structure, demonstrating that the number of bins required to capture a given fraction of the total power is insufficient to fully characterize the differences between these trajectories. The bin counts therefore provide a useful measure of spectral concentration, but must be considered alongside the spectral profiles themselves to distinguish between different forms of spectral structure. The rapid increase in the number of bins required to capture the final few percent of the total power indicates that substantial power is distributed among numerous low-amplitude spectral components. This is consistent with the harmonic decomposition commonly used to interpret EMRI waveforms as a superposition of distinct harmonic contributions, often described as a collection of "voices" each associated with a particular orbital harmonic \citep{Hughes2001, Speri_et_al_2024}.

\begin{table}
\centering
\caption{Dominant spectral peaks of the representative trajectories.}
\begin{tabular}{lccc}
\hline
Orbit & $f$ ($1/M$) & Best Mode $(m,k,n)$
     & Power (\%)\\
\hline
Quasi-circular & $1.0210\times10^{-2}$ & $(2,0,0)$ & $99.87$\\
\hline
Zoom--Whirl & $1.2275\times10^{-1}$ & $(2^*,-2,1)$ & $5.21$\\
            & $1.2371\times10^{-1}$ & $(2^*,-2,2)$ & $5.14$\\
            & $1.2179\times10^{-1}$ & $(2^*,-2,0)$ & $5.10$\\
\hline
Generic & $8.7134\times10^{-3}$ & $(2,0,1)$ & $21.82$\\
        & $4.3545\times10^{-3}$ & $(2,0,-1)$ & $13.00$\\
        & $1.0893\times10^{-2}$ & $(2,0,2)$ & $12.07$\\
        
\hline
Near-polar  & $1.4094\times10^{-2}$ & $(2,0,2)$ & $12.51$\\
            & $5.9724\times10^{-3}$ & $(2,0,-1)$ & $11.26$\\
            & $1.1387\times10^{-2}$ & $(2,0,1)$ & $9.15$\\
\hline
\multicolumn{4}{l}{\footnotesize $^*$ Periapsis-associated frequency used for zoom--whirl modes.}
\end{tabular}
\label{tab:geodesic_spectral_content}
\end{table}

The dominant peaks identified in Table~\ref{tab:geodesic_spectral_content} are broadly consistent with the expected harmonic structure. In the quasi-circular case, the dominant peak corresponds to the $(m,k,n)=(2,0,0)$ harmonic, as expected for a nearly circular orbit whose radiation is dominated by azimuthal motion \citep{Hughes2001}. The generic trajectory exhibits several strong radial sidebands of the dominant azimuthal harmonic. The most prominent modes $(2,0,1)$, $(2,0,2)$, and $(2,0,-1)$ correspond to frequencies $2f_\phi + f_r$, $2f_\phi + 2f_r$, and $2f_\phi - f_r$, respectively, and account for a substantial fraction of the total power. The near-polar trajectory exhibits a richer combination of radial and polar harmonics, with several distinct $(m,k,n)$ combinations contributing comparably to the dominant spectral content.

The Zoom--Whirl trajectory requires somewhat different interpretation. Matching its high-frequency peaks using the global fundamental frequencies produces substantially poorer agreement than for the other trajectories. However, replacing the global azimuthal/radial description with the frequency associated with the rapid motion near periapsis produces much closer matches for the dominant peaks; this behavior is consistent with other analyses of Zoom--Whirl motion, where the spectrum is often similar to that of a circular orbit of the same angular velocity at periapsis, along with many harmonics of that frequency \citep{Glampedakis_Kennefick_2002, Burko_Khanna_2007,Enoki_2007,Tessmer_2007}. The rapid phase accumulation during the 'whirl' portion of the orbit introduces strong high-frequency structure that is not well represented by treating the entire orbit as a single characteristic azimuthal cycle. The resulting spectrum therefore is consistent with the numerical trajectory resolving the short-timescale structure of the near-separatrix orbit.

\begin{figure}
    \centering
    \includegraphics[scale=0.4]{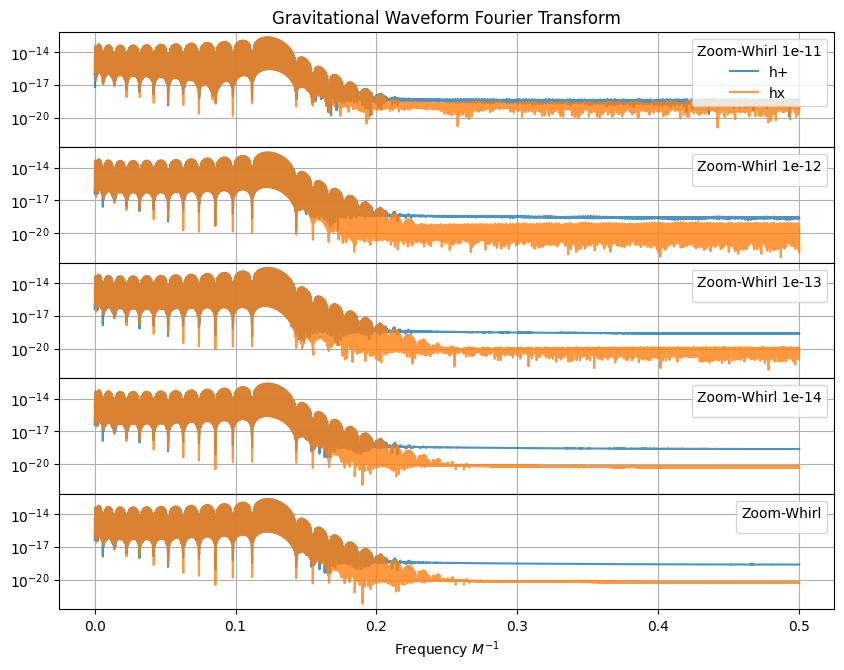}
    \caption{Fourier amplitudes of the quadrupole-based waveform for the Zoom--Whirl trajectory at integration tolerances ranging from $10^{-15}$ to $10^{-11}$. The dominant spectral features remain stable as the tolerance is varied, while high-frequency noise eventually decreases with strict enough constraints.}
    \label{fig:geodesic_spectral_convergence}
\end{figure}

As an additional numerical check, the spectral content of the representative trajectories was compared across several integration tolerances. The dominant features remain stable as the tolerance is relaxed, indicating that they are not artifacts of the particular integration accuracy used for the baseline trajectories. Differences become increasingly apparent toward the highest sampled frequencies. This behavior is particularly clear for the generic trajectory, for which progressively looser tolerances obscure the fine clustered structure at high frequencies while leaving the dominant low-frequency features essentially unchanged.

This high-frequency sensitivity is expected because the proxy waveform requires both interpolation onto a uniform time grid and numerical differentiation of the quadrupole moment. Small-scale numerical errors in the trajectory can therefore be amplified in the second time derivative and subsequently appear preferentially at high frequencies. We consequently regard the dominant and well-converged spectral features as the most useful diagnostic of the numerical trajectories, while treating fine structure near the upper end of the sampled frequency range with increasing caution.

Overall, the Fourier analysis demonstrates that the representative geodesics reproduce the expected qualitative spectral structure of bound Kerr motion. Nearly circular motion produces a comparatively concentrated spectrum, while eccentricity, inclination, and strong-field Zoom--Whirl motion introduce increasingly rich combinations of radial, polar, and azimuthal harmonics. The stability of the dominant features under changes in integration tolerance provides an additional validation of the numerical trajectories and of the subsequent waveform construction.

\subsection{Radiation Reaction Selection}
\label{sec:rrs}

While geodesic tests establish the accuracy of the underlying orbital integration, the inclusion of radiation reaction introduces a separate source of modeling uncertainty: the prescription used to determine the evolution of the Carter constant. We therefore compare the three implementations described above and select a prescription for use in the remainder of the study.

We first compare the prescriptions over a broad range of orbital parameters by evaluating the average rates of change of the orbital parameters over 10 radial orbits. The dimensionless black-hole spin is varied over $a\in \{0.1,0.5,0.9\}$, while the eccentricity is varied over $e\in \{0.01,0.5,0.9\}$. For each $(a,e)$ combination, four values of the semilatus rectum are considered. We use $p=60$ as a representative weak-field orbit and $p=30$ as an intermediate-field case. Two additional values probe the strong-field regime: one is set to $1.1$ times the minimum semilatus rectum for a retrograde orbit of the specified eccentricity, while the other is set to $1.1$ times the corresponding minimum for a prograde orbit. In all cases, we set the mass ratio $\mu/M = 10^{5}$.

The latter choice provides a particularly strong-field test for rapidly spinning black holes. Because the prograde separatrix moves inward with increasing spin, the resulting value of $p$ can correspond to an increasingly large range of inclinations for which the orbit is no longer physically allowed. Evaluating the prescriptions at fixed $p$ nevertheless provides a useful diagnostic of their behavior as the boundary of the allowed orbital parameter space is approached.

\begin{figure}
\centering
\includegraphics[scale=0.45]{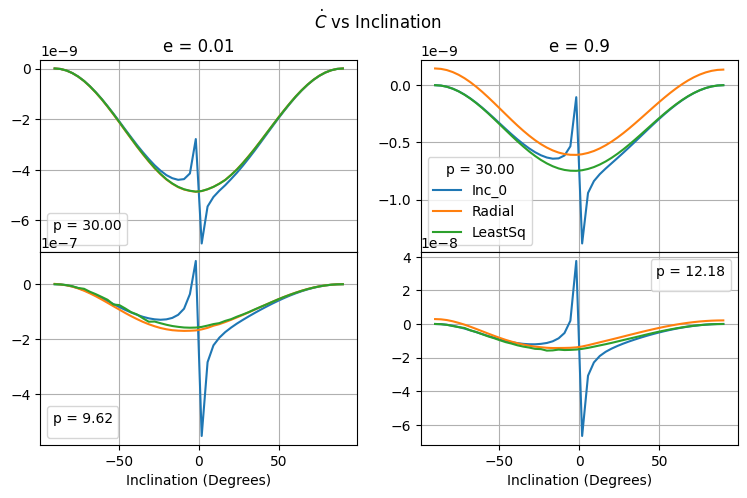}
\caption{Average Carter-constant evolution rate as a function of inclination for representative points in the parameter-space survey. The plotted cases correspond to $a=0.9$, $e=0.01$ and $0.9$, and the intermediate- and strong-field values of $p$.}
\label{fig:cdot_spread}
\end{figure}

\begin{figure}
\centering
\includegraphics[scale=0.45]{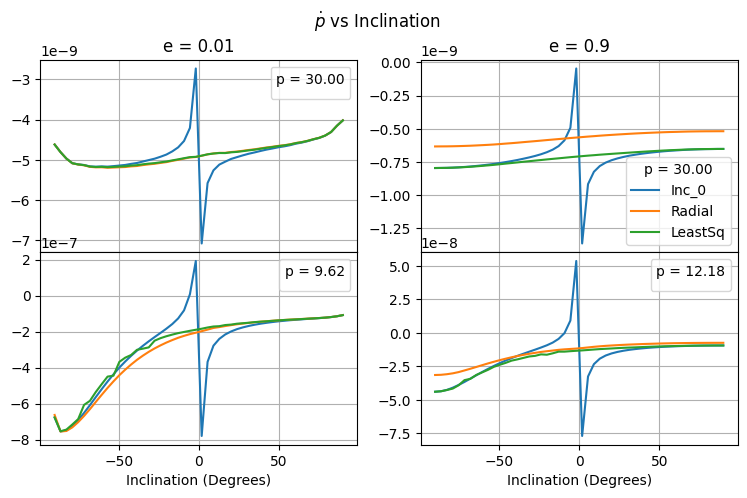}
\caption{Average semilatus-rectum evolution rate for the same representative parameter combinations shown in Figure~\ref{fig:cdot_spread}.}
\label{fig:pdot_spread}
\end{figure}

The three prescriptions exhibit broadly similar inclination dependence away from their pathological regions, providing a basic consistency check on the calculation. Figures~\ref{fig:cdot_spread} and~\ref{fig:pdot_spread} show representative results for the high-spin case, $a=0.9$, at low and high eccentricity and for distinct values of $p$. The full parameter survey exhibits the same general qualitative behavior across the sampled values of spin, eccentricity, and semilatus rectum. The dependence on inclination becomes increasingly asymmetric between prograde and retrograde motion as the spin increases, while the characteristic midpoint of the curves changes comparatively little. Decreasing $p$ generally increases the magnitude of the radiation-reaction-driven changes and enhances the inclination dependence, consistent with the increasing importance of strong-field effects. Conversely, increasing eccentricity generally reduces the magnitude of the rates at fixed $p$, consistent with the larger characteristic radial scale of the corresponding orbit.

Despite these broad similarities, several systematic differences are apparent between the prescriptions. The constant-inclination prescription develops a pronounced discontinuity near polar orbits in both $\dot C$ and $\dot p$. In some cases, this produces positive values of $\dot C$ and $\dot p$ for Near-Polar retrograde trajectories, implying an increase in quantities that are otherwise expected to decrease under radiation reaction, or extremely large excursions in the opposite direction, implying an unphysically rapid evolution. This behavior is sufficiently pathological to rule out the constant-inclination prescription in its present implementation.

The radial prescription avoids this discontinuity, but exhibits a small systematic bias in $\dot C$ that becomes increasingly apparent at higher eccentricity. In particular, the prescription can produce small positive values of $\dot C$ for planar and nearly planar orbits. The effect becomes particularly apparent in the weak-field portion of the parameter survey, where the physical rates are themselves small. We found that this behavior is strongly connected to the treatment of the radial contribution in the differential evolution of the Carter constant. In particular, evaluating the radial term at different locations along the orbit can substantially alter both the magnitude and sign of the resulting bias. The final expression adopted for the radial prescription, Eq.~\eqref{eq:[radial-method-equation]}, was selected because it minimizes this systematic offset among the alternatives investigated.

At this stage, the least-squares prescription appears to be the most robust of the three methods. It has no comparable polar discontinuity, exhibits little systematic bias in $\dot C$, and maintains qualitatively consistent behavior across the parameter ranges examined. Based solely on the instantaneous parameter-space comparison, the least-squares prescription therefore appears to provide the most favorable behavior. However, instantaneous behavior does not provide a complete test of the radiation reaction prescriptions. This motivates a comparison of their behavior over complete inspirals using the four representative trajectories introduced previously.

\begin{figure*}[t]
    \centering
    \subfloat[]{
        \includegraphics[scale=0.57]{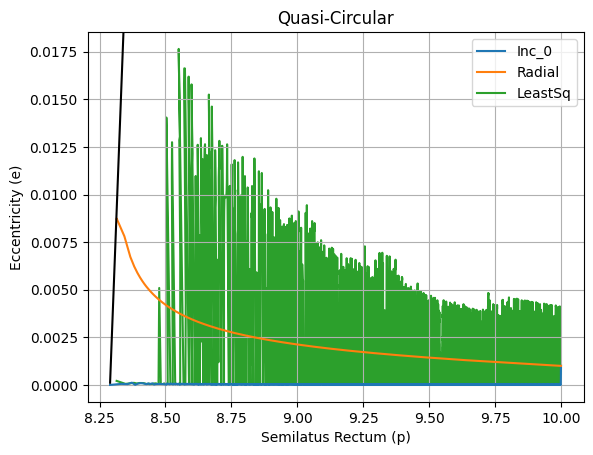}
        \label{fig:rr_quasicircular_pve}
    }
    \hfill
    \subfloat[]{
        \includegraphics[scale=0.57]{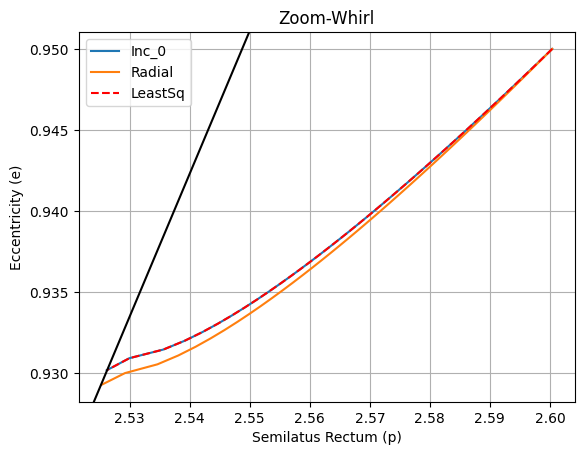}
        \label{fig:rr_zoomwhirl_pve}
    }
    \vspace{0.5em}
    \subfloat[]{
        \includegraphics[scale=0.57]{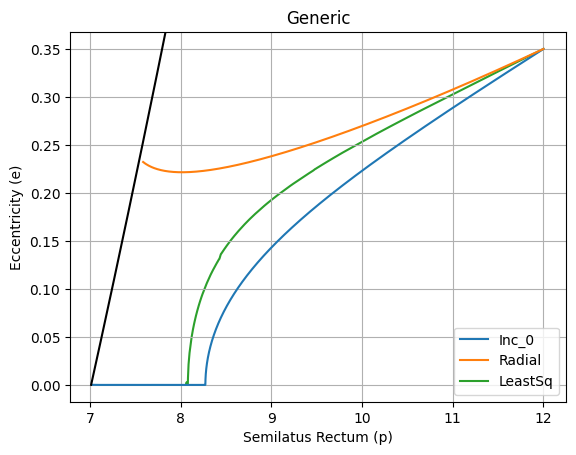}
        \label{fig:rr_generic_pve}
    }
    \hfill
    \subfloat[]{
        \includegraphics[scale=0.57]{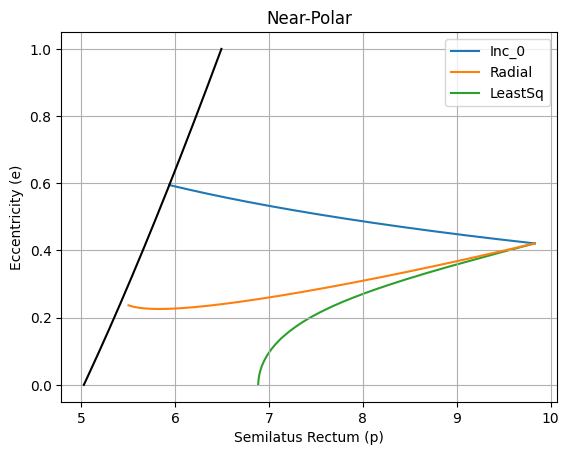}
        \label{fig:rr_nearpolar_pve}
    }
    \caption{Evolution of the representative trajectories in the $(p,e)$ plane for the three radiation reaction prescriptions. In each figure, the black line is the separatrix between bound and unbound orbits for the given dimensionless spin. Note that the Least Squares prescription in Fig.~\ref{fig:rr_quasicircular_pve} exhibits extreme variations in eccentricity, while the constant-inclination prescription drops to $e=0$ immediately.}
    \label{fig:rr_representative_pve}
\end{figure*}

The resulting $(p,e)$ trajectories reveal substantial differences between the prescriptions. In contrast to the parameter-space comparison, the least-squares prescription frequently produces qualitatively implausible long-term behavior. Figure~\ref{fig:rr_representative_pve} shows the resulting evolution for the four representative trajectories. In the Generic and and Near-Polar trajectories, the eccentricity decreases rapidly toward $e=0$ as the system evolves; this behavior is inconsistent with the expected evolution, where the rate of circularization generally decreases or even reverses \citep{Ryan_1996, Kennefick_1998, Glampedakis_Kennefick_2002}.  The circularization of these orbits accelerate with time against predictions, indicating that the associated increase in periapsis is not an expected behavior under normal radiation-reaction driven evolution. The Quasi-Circular case exhibits similarly strong sensitivity, with the least-squares prescription producing substantial fluctuations in eccentricity during the inspiral, in contrast to the comparatively smooth evolution produced by the radial prescription.

The constant-inclination prescription also exhibits problematic behavior, including rapid circularization in the Generic and Quasi-Circular cases. The Zoom--Whirl trajectory is the least discriminating test in this regard, as all three prescriptions produce broadly similar $(p,e)$ evolution over the relatively short portion of the inspiral considered.

In contrast, the radial prescription produces consistently smooth orbital evolution across the representative trajectories. In particular, the eccentricity evolution remains gradual and does not exhibit the abrupt collapse toward circularity seen in the least-squares and constant-inclination prescriptions. For the Generic and Near-Polar trajectories, the radial prescription also exhibits a turnaround in eccentricity evolution as the orbit approaches the strong field, rather than continuing toward an artificially circular state.

Another useful additional diagnostic is the evolution of the orbital inclination. Because the energy and axial angular-momentum fluxes are common to all three prescriptions, differences in the inclination evolution provide a complementary diagnostic of the consequences of the different Carter-constant prescriptions. An inappropriate $\dot C$ can consequently produce an angular-momentum evolution that is inconsistent with the energy loss.

\begin{figure*}[t]
    \centering
    \subfloat[]{
        \includegraphics[scale=0.57]{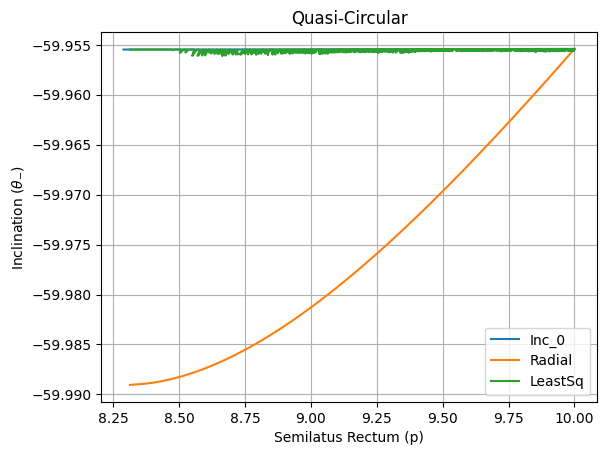}
        \label{fig:rr_quasicircular_pvi}
    }
    \hfill
    \subfloat[]{
        \includegraphics[scale=0.57]{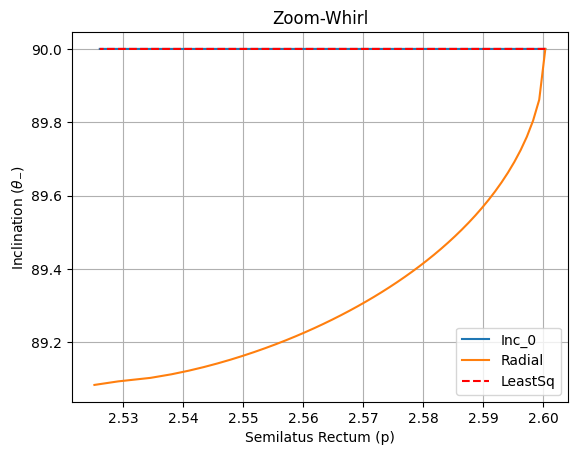}
        \label{fig:rr_zoomwhirl_pvi}
    }
    \vspace{0.5em}
    \subfloat[]{
        \includegraphics[scale=0.57]{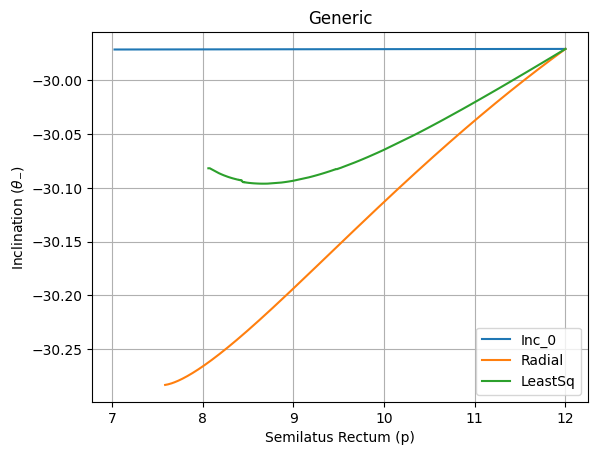}
        \label{fig:rr_generic_pvi}
    }
    \hfill
    \subfloat[]{
        \includegraphics[scale=0.57]{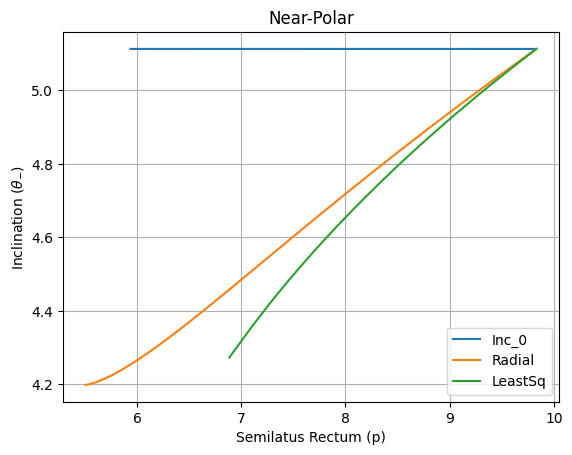}
        \label{fig:rr_nearpolar_pvi}
    }
    \caption{Evolution of the inclination of the representative trajectories for the three radiation reaction prescriptions.}
    \label{fig:rr_representative_pvi}
\end{figure*}

In the Newtonian limit, this behavior can be understood qualitatively from the fact that circular orbits have the maximum angular momentum available at fixed orbital energy. Thus, if the angular-momentum content of an orbit is reduced too slowly relative to its energy, the orbit is driven toward circularity. For generic Kerr orbits, the Carter constant contributes to the total angular-momentum structure, so this argument should be regarded as qualitative rather than as a direct criterion on $\dot C$ alone. Nevertheless, the observed correlation between anomalous $\dot C$ prescriptions and rapid circularization provides evidence that the long-term failures arise from the coupled evolution of the orbital constants rather than from the underlying geodesic integrator.

The inclination evolution also distinguishes the long-term behavior of the prescriptions. As expected, the constant-inclination prescription maintains an essentially fixed inclination, with deviations from the initial value of no more than approximately $10^{-6}$ degrees in the representative cases. The least-squares prescription exhibits substantially more varied behavior. For the Generic trajectory, its inclination initially evolves between the behaviors of the constant-inclination and radial prescriptions, but begins moving toward more prograde values at approximately the same point at which the periapsis begins to increase. This coincides with the onset of the unphysical long-term evolution discussed above. For the Near-Polar trajectory, the least-squares prescription produces the most rapid deviation from the initial inclination, evolving from approximately $5.05^\circ$ from the polar orientation to roughly $4.0^\circ$ before catastrophic circularization. In the Quasi-Circular case, the inclination undergoes small fluctuations between slightly more prograde and more retrograde values before settling approximately at its initial value. By contrast, the Zoom--Whirl trajectory remains nearly constant in inclination under the least-squares prescription.

The radial prescription exhibits comparatively consistent inclination evolution across the representative trajectories. The Generic orbit evolves from approximately $-29.98^\circ$ to $-30.28^\circ$, while the Near-Polar orbit changes from approximately $5.05^\circ$ to $4.2^\circ$ from the polar orientation. The Quasi-Circular orbit undergoes only a small shift toward more retrograde inclination. The principal exception is the equatorial Zoom--Whirl trajectory, for which the radial prescription produces substantially more rapid changes in both inclination and eccentricity than the other two prescriptions. This behavior is consistent with the instantaneous parameter-space results, which show that the radial prescription develops a positive $\dot C$ for planar and nearly planar orbits in some regions of parameter space. The discrepancy therefore appears to be associated with the same near-planar behavior offset in $\dot{C}$ identified in the broader parameter survey rather than with a general instability of the radial prescription.

Taken together, these results demonstrate that local agreement between radiation-reaction prescriptions is not sufficient to predict their long-term behavior. The parameter-space survey is useful for identifying systematic differences and pathological regions, but the full inspiral integrations provide the more stringent test because small differences in the instantaneous orbital fluxes can accumulate into qualitatively different trajectories. In particular, the least-squares prescription, despite appearing favorable in the instantaneous comparison, repeatedly produces rapid circularization and associated anomalous orbital evolution in the representative inspirals. The radial-potential prescription is the only method among the prescriptions tested here that consistently avoids these failures across the full set of representative trajectories. We therefore select the radial-potential prescription for the remainder of this work.

\subsection{Extended Validation with Radiation Reaction}
\label{RR_Validation}

Having selected the radial-potential prescription for the evolution of the Carter constant, we next assess the numerical behavior of inspirals with radiation reaction enabled. These tests use a mass ratio of $\mu=10^{-5}$, in line with the radiation reaction selection tests described above. The purpose of this section is not to repeat the geodesic validation, but rather to determine whether the inclusion of radiation reaction introduces additional numerical sensitivity and whether the previously identified spectral and convergence properties persist during inspiral. For ease of comparison, each representative trajectory that is run for up to 500 radial orbits, as in Sec.~\ref{sec:geo_val}. The exception is the Zoom--Whirl trajectory, which plunges after 48 radial orbits. This early plunge is consistent with the trajectory's proximity to the bound--plunge separatrix, where relatively small changes in the orbital parameters can lead to plunge.

The coordinate-space trajectories provide relatively little additional information in this comparison and are therefore not considered in detail. For the representative trajectories, radiation reaction produces only modest changes over the intervals considered, with the most visible differences occurring for the Zoom--Whirl orbit. Instead, we focus on quantities that provide more direct tests of numerical convergence and the preservation of the orbital frequency structure during inspiral.

\begin{figure}
\centering
\includegraphics[scale=0.57]{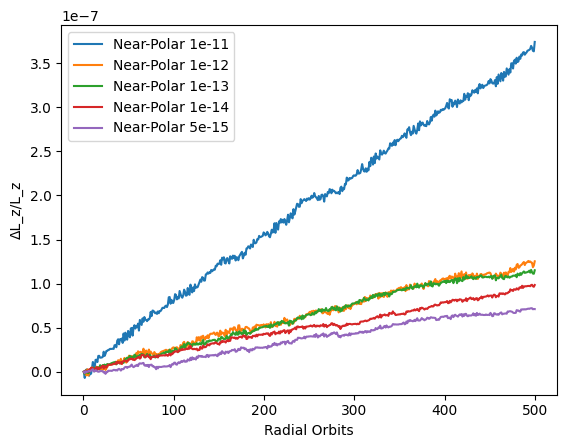}
\caption{Fractional deviation in $L_z$ between integrations performed at looser tolerances and the reference calculation with an integration tolerance of $10^{-15}$ for the Near-Polar trajectory. Across the representative trajectories, the $10^{-11}$ calculation generally exhibits the largest deviation, while tighter tolerances remain substantially closer to the reference solution.}
\label{fig:rr_convergence}
\end{figure}

As an initial convergence test, the orbital evolution was compared for several integration tolerances. Because the orbital constants undergo evolution under radiation reaction, their convergence cannot be assessed by comparison with their initial values as in the geodesic calculations. Instead, the results from the looser integration tolerances are compared with the calculation using an integration tolerance of $10^{-15}$, which serves as the reference solution. Because the orbital constants are updated once per radial orbit, changes in integration tolerance are not expected to directly alter the sampling of the orbital constants; however, to account for small differences in the timing of the radial-orbit updates between integrations, the results from the looser tolerances are interpolated onto the sampling times of the reference calculation before computing the fractional deviations. A representative example is shown in Fig.~\ref{fig:rr_convergence}.

The resulting fractional deviations show general convergence toward the $10^{-15}$ reference calculation as the integration tolerance is decreased, although the convergence is not uniformly linear over the full evolution. Small differences in the numerically integrated trajectory can accumulate over successive radiation-reaction updates, producing departures from a simple tolerance-scaling relation, but this does not indicate a significant change in the inspiral dynamics; in other words, {\it the true error is dominated by the radiation reaction prescription as opposed to the geodesic integration.}

\begin{figure}
\centering
\includegraphics[scale=0.55]{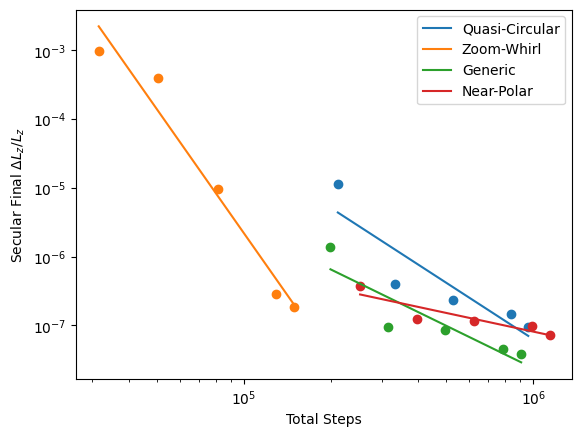}
\caption{Fractional deviation in the final value of $L_z$ from the $10^{-15}$ reference calculation as a function of the total number of integration steps for the representative trajectories, as a counterpoint to Fig.~\ref{fig:tolerance_convergence} with radiation reaction activated. Although the individual results do not follow a single linear convergence relation, integrations performed at tighter tolerances generally produce smaller deviations from the reference solution. The Zoom--Whirl has the greatest divergence from the general trend, a consequence of its much more rapid plunge.}
\label{fig:secular_lz_convergence_rr}
\end{figure}

\begin{figure}
\centering
\includegraphics[scale=0.49]{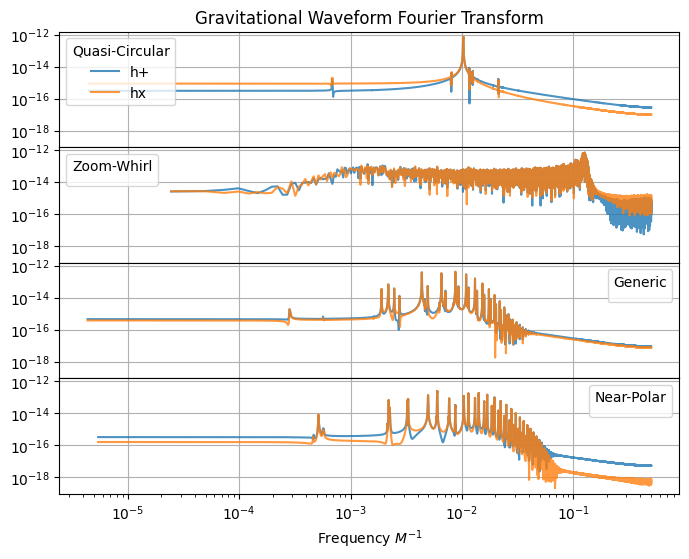}
\caption{Fourier transforms of the representative trajectories with radiation reaction enabled. The underlying harmonic structure remains consistent with that of the corresponding geodesic trajectories, while evolution of the orbital frequencies during the inspiral distributes power over broader frequency ranges. This is most apparent in the Zoom--Whirl spectrum.}
\label{fig:rr_spectra}
\end{figure}

The Fourier spectra provide an additional test of how radiation reaction alters the orbital frequency structure over the timescale considered. Figure~\ref{fig:rr_spectra} shows the Fourier transforms of the four representative trajectories with radiation reaction enabled. The overall spectral structure remains consistent with that obtained from the corresponding geodesic trajectories, but the evolution of the orbital parameters during the integration produces broadening of the spectral features as different harmonics are strengthened and peaks migrate. Rather than remaining localized at a single frequency, power is distributed over a range of frequencies as the characteristic orbital frequencies evolve throughout the inspiral, as is illustrated in Table \ref{tab:char_freq_changes}. This is most evident in the Zoom-Whirl spectrum, as the 'smearing' of frequencies largely disrupts the complex structure visible in the geodesic example. 

\begin{table}
\centering
\caption{Change in characteristic frequencies}
\begin{tabular}{lccc}
\hline
Orbit & $\Delta f_{\phi}$ & $\Delta f_{\theta}$ & $\Delta f_r$\\
\hline
Quasi-Circular & $7.448\times 10^{-5}$ & $8.448\times 10^{-5}$ & $-1.973\times 10^{-5}$\\
Zoom--Whirl & ${9.105\times 10^{-3}}^*$ & $2.745\times 10^{-3}$ & $3.711\times 10^{-4}$\\
Generic & $1.323\times 10^{-5}$ & $1.424\times 10^{-5}$ & $5.898\times 10^{-6}$\\
Near-Polar & $2.968\times 10^{-5}$ & $2.694\times 10^{-5}$ & $1.330\times 10^{-5}$\\
\hline
\multicolumn{4}{l}{\scriptsize $^*$The periapse-associated azimuthal frequency, $\Delta f_{\phi p} = -5.799\times 10^{-5}$}
\end{tabular}
\label{tab:char_freq_changes}
\end{table}

The magnitude of this effect can be observed more clearly by examining changes in the fundamental frequencies of the representative trajectories, which determine the locations of the corresponding orbital harmonics. As shown in Table~\ref{tab:char_freq_changes}, the Zoom--Whirl trajectory exhibits the largest changes by a substantial margin: its azimuthal, polar, and radial frequencies change by $9.105\times10^{-3}$, $2.745\times10^{-3}$, and $3.711\times10^{-4} M^{-1}$, respectively. Each of these changes is comparable in scale to its corresponding initial frequency in Table~\ref{tab:char_freq_inits}. This pronounced evolution is consistent with the trajectory's proximity to the bound--plunge separatrix, near which radiation-reaction-driven evolution or orbital parameters, and thus the fundamental frequencies, begins to accelerate \citep{Glampedakis_Kennefick_2002, Schmidt2002}. By comparison, the frequency changes for the Generic and Near-Polar trajectories are no greater than a few times $10^{-5}M^{-1}$, while those for the Quasi-Circular trajectory are of a similar order. One notable exception to the general increase in frequency is the radial frequency of the Quasi-Circular trajectory, which decreases by $1.973\times10^{-5} M^{-1}$. This behavior is consistent with the non-monotonic radial epicyclic frequency of circular Kerr orbits, which reaches a maximum outside the innermost stable circular orbit (ISCO) before decreasing toward zero as the ISCO is approached \citep{Schmidt2002, Torok_Stuchlik_2005}. Finally, the periapsis-associated azimuthal frequency of the Zoom--Whirl trajectory changes by only $-5.799\times10^{-5} M^{-1}$, remaining nearly constant despite the substantial changes in the three fundamental frequencies.

The broadening of the spectra is also reflected quantitatively in the distribution of spectral power. Table~\ref{tab:geodesic_spectral_binning_rr} lists the number of frequency bins required to contain 50\%, 90\%, and 99\% of the total power for each representative trajectory with radiation reaction enabled. Compared with the corresponding geodesic spectra as listed in Table~\ref{tab:geodesic_spectral_binning}, all four trajectories require more bins to capture 50\% and 90\% of the total power, although the extent of this change varies considerably. The most pronounced broadening occurs for the Zoom--Whirl trajectory, for which the number of bins required to capture 99\% of the power increases from 1106 to 4181. The Quasi-Circular spectrum also broadens substantially, from a single bin containing 99\% of the total power to 53 bins. The Generic spectrum shows a more modest increase, from 206 to 312 bins at the same threshold. For the Near-Polar trajectory, however, the number of bins required to capture 99\% of the power decreases slightly, from 459 to 429, despite the increase at the 50\% and 90\% thresholds. This may reflect a modest reduction in the contribution of the weakest, higher-frequency components, although the bin counts alone do not establish the cause. These results show that the redistribution of spectral power is not uniform across all trajectories or frequency ranges.

\begin{table}
\centering
\caption{Fourier Spectral Binning Power Distribution \\(Radiation Reaction Enabled)}
\begin{tabular}{lccc}
\hline
Orbit & 50\% & 90\% & 99\%  \\
\hline
Quasi-circular & 15 & 30 & 53 \\
Zoom--whirl    & 204 & 1676 & 4181 \\
Generic        & 12 & 48 & 312 \\
Near-polar     & 36 & 144 & 429\\
\hline
\end{tabular}
\label{tab:geodesic_spectral_binning_rr}
\end{table}

The redistribution is also evident in the dominant spectral peaks. In the geodesic spectra, the strongest individual peaks account for approximately 13-22\% of the total power for the Generic and Near-Polar trajectories, whereas the strongest peaks in the corresponding RR-enabled spectra account for only about 3-6\%. The strongest peak in the Zoom--Whirl spectrum similarly decreases from approximately 5\% to 1\%, while the Quasi-Circular spectrum changes from a single dominant peak containing nearly all of the power to several nearby peaks, each containing no more than 4\% of the total power. Because radiation reaction causes the orbital frequencies to evolve throughout the inspiral, these differences reflect both the redistribution of power among spectral components and the broadening and shifting of their frequencies. The underlying harmonic structure remains recognizable, but the full inspiral spectrum is less concentrated in a small number of sharply localized peaks.

Taken together, these results show that radiation reaction produces measurable spectral evolution over the timescales considered, but does not qualitatively alter the underlying frequency structure of the Kerr geodesics. The dominant harmonic content remains recognizable, while secular evolution of the orbital parameters produces the expected broadening of the corresponding spectral features.

\subsection{Comparison with Existing Radiation-Reaction Prescriptions}
\label{RR_Comparison}

We next compare the selected radial-potential prescription with several existing approximations for radiation reaction. We first consider the orbit-averaged results of ~\citet{Peters1964}, specifically the treatment of two point masses on Keplerian elliptical orbits developed in Section V. This provides a weak-field, Newtonian baseline that does not account for black-hole spin or other relativistic effects. We then consider the prescription of Glampedakis, Hughes, and Kennefick~\citep{Glampedakis2002}, hereafter GHK, which applies a constant-inclination prescription similar to the one described in this paper, and that of Gair and Glampedakis~\citep{Gair2006}, hereafter GG, which extends the GHK treatment with modifications intended to improve its applicability to low-eccentricity and high-inclination orbits. These comparisons provide an external check on the qualitative behavior of the implementation while illustrating how the selected prescription relates to approximations of increasing physical complexity. 

\begin{figure*}[t]
    \centering
    \subfloat[]{
        \includegraphics[scale=0.57]{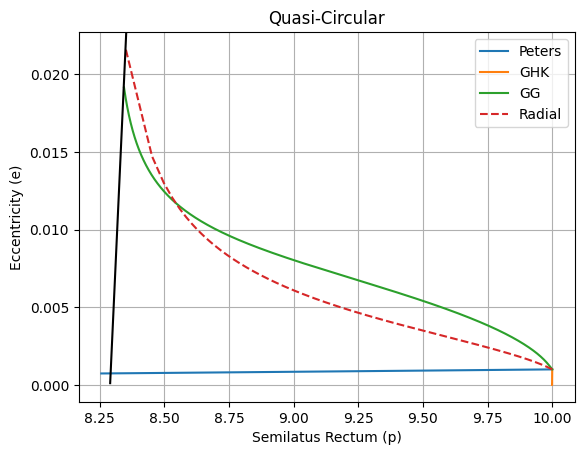}
        \label{fig:comps_circle_pve}
    }
    \hfill
    \subfloat[]{
        \includegraphics[scale=0.57]{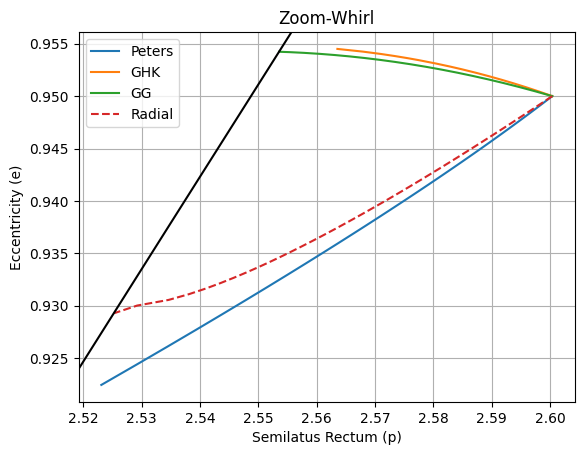}
        \label{fig:comps_zoom_pve}
    }

    \vspace{0.5em}

    \subfloat[]{
        \includegraphics[scale=0.57]{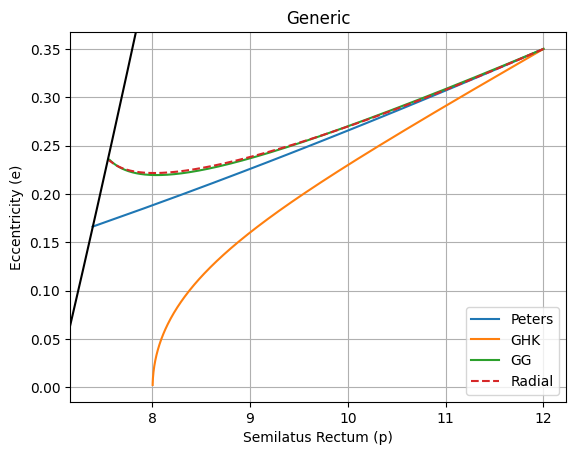}
        \label{fig:comps_generic_pve}
    }
    \hfill
    \subfloat[]{
        \includegraphics[scale=0.57]{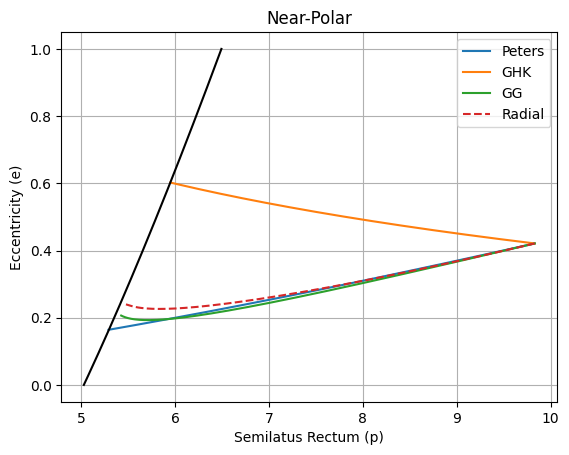}
        \label{fig:comps_polar_pve}
    }

    \caption{Evolution of the representative trajectories in the $(p,e)$ plane for the `Radial' radiation reaction prescription and three approximations.}
    \label{fig:comps_representative_pve}
\end{figure*}

The radial-potential prescription shows similar secular evolution to the GG method across the representative inspirals, with the closest correspondence occurring for the Generic and Near-Polar trajectories. This agreement is particularly useful because the GG prescription incorporates a more sophisticated treatment of the evolution of inclined and eccentric Kerr orbits than the simpler approximations considered here. The close correspondence therefore provides an independent indication that the selected prescription captures the expected secular orbital evolution.

\begin{figure*}[t]
    \centering
    \subfloat[]{
        \includegraphics[scale=0.57]{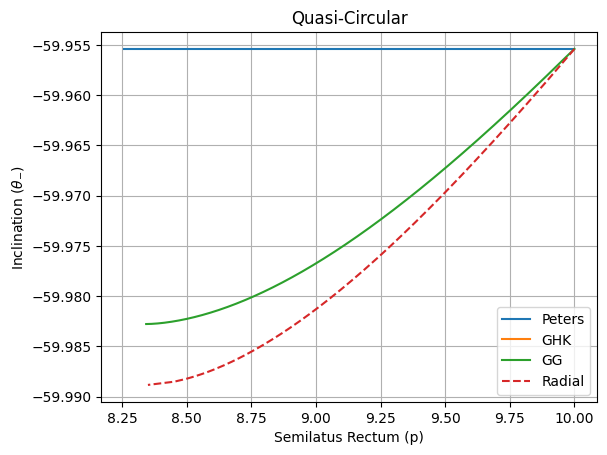}
        \label{fig:comps_circle_pvi}
    }
    \hfill
    \subfloat[]{
        \includegraphics[scale=0.57]{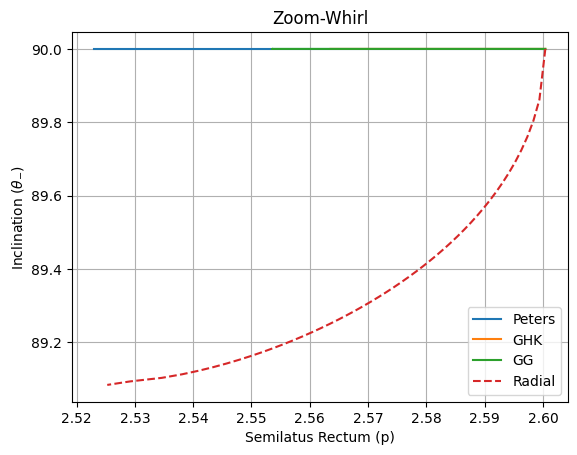}
        \label{fig:comps_zoom_pvi}
    }
    \vspace{0.5em}
    \subfloat[]{
        \includegraphics[scale=0.57]{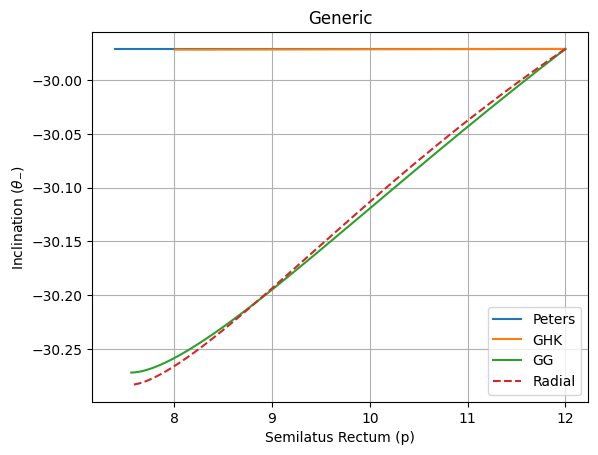}
        \label{fig:comps_generic_pvi}
    }
    \hfill
    \subfloat[]{
        \includegraphics[scale=0.57]{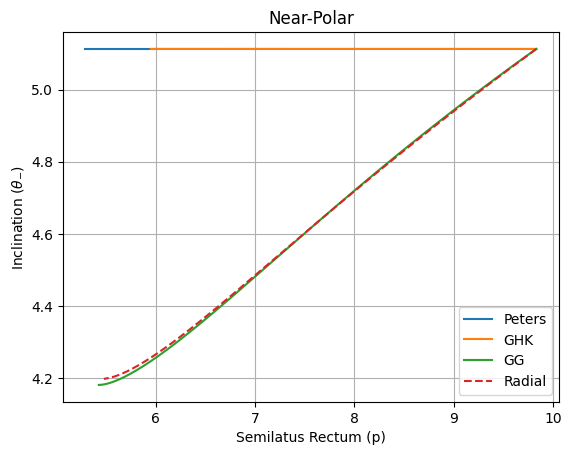}
        \label{fig:comps_polar_pvi}
    }
    \caption{Evolution of the inclination of the representative trajectories for the Radial radiation reaction prescription and three approximations.}
    \label{fig:comps_representative_pvi}
\end{figure*}

The largest differences occur for the Zoom--Whirl and Quasi-Circular trajectories. In the Zoom--Whirl case, the radial prescription exhibits a somewhat different $(p,e)$ evolution from GG, accompanied by the previously identified difference in the evolution of the Carter constant for nearly planar orbits. In the Quasi-Circular case, the orbital inclinations remain broadly consistent between the two prescriptions, while GG produces a more aggressive initial growth in eccentricity and a smaller overall change in inclination than the radial prescription. The origin of this discrepancy is not established here, as resolving it would require a more detailed comparison of the assumptions and flux prescriptions used by the two methods. 

At the beginning of the integrations, the GG prescription also generally agrees well with the Peters prescription for the representative non-Zoom--Whirl trajectories. This agreement is expected because the Peters prescription describes the leading-order radiation reaction for Keplerian binaries and should therefore provide a reasonable weak-field reference. The correspondence is less useful for the Zoom--Whirl trajectory, whose strong-field dynamics lie outside the assumptions underlying the Peters approximation.

The GHK prescription shows a similar qualitative evolution to the constant-inclination prescription considered earlier. This correspondence is consistent with the fact that both prescriptions impose a fixed-inclination approximation in their treatment of the Carter constant. Although this agreement does not provide an independent validation of the radial-potential prescription, it does provide a useful consistency check on the interpretation of the earlier parameter-space results.

Taken together, the comparisons with existing prescriptions provide an external consistency check on the radial-potential treatment. Its close agreement with GG across most of the representative trajectories is consistent with the smooth long-term evolution observed in the preceding tests, while the differences identified in the Quasi-Circular and Zoom--Whirl cases highlight regimes in which the underlying approximations produce measurably different behavior. These comparisons do not establish the radial-potential prescription as uniquely correct, but they demonstrate that its behavior is broadly consistent with established radiation-reaction treatments while avoiding the pathological evolution observed for the alternative prescriptions considered here.

\section{Discussion and Future Applications}
\label{sec:discussion}

The results presented above demonstrate that \EMRICode{}can be used to evolve generic Kerr geodesics under a prescribed radiation-reaction model while retaining direct control over the orbital dynamics and the prescription used to modify the orbital constants of motion. The modular structure of the implementation also allows the radiation-reaction prescription to be modified independently of the underlying geodesic evolution. This provides a natural framework for testing alternative prescriptions and for extending the model beyond the approximations considered here. Several aspects of the present implementation, however, remain approximate and provide opportunities for future development.

\subsection{Limitations of the Current Implementation}

The principal limitation of the current radiation-reaction implementation is the prescription used to determine the evolution of the Carter constant. As discussed in Sec.~\ref{sec:rrs}, the radial-potential prescription considered here performs substantially better than the alternative prescriptions tested in this work, but it nevertheless exhibits a systematic residual bias. In particular, for nearly planar orbits the prescription can produce a positive evolution of $C$ in regimes where the expected evolution is non-positive, causing initially equatorial or nearly equatorial trajectories to rapidly acquire inclination when none is expected. The results suggest that at least part of this discrepancy is associated with the radius at which the radial potential is evaluated. Averaging the prescription over the radial motion, as in Eq.~\ref{eq:[radial-method-equation]}, substantially reduces the effect, suggesting that a more appropriate radial weighting may provide a path toward a more generally accurate prescription.

A useful direction for future work would therefore be to investigate the origin of this residual bias rather than treating it solely as an empirical correction. In particular, the dependence of the discrepancy on orbital radius, eccentricity, and inclination could be used to determine whether the bias arises from the local evaluation of the radial potential, from the mapping between instantaneous orbital quantities and orbit-averaged fluxes, or from the underlying approximation used to relate changes in the constants of motion. A prescription based on an appropriately weighted average over the radial cycle could potentially retain the simplicity of the current method while reducing its systematic error. 

The least-squares prescription considered in this work presents a separate limitation. Although it provides a formally straightforward way of relating changes in the conserved quantities to changes in the local four-velocity, in practice it produced substantially less reliable long-term evolution for the systems tested here. In particular, some highly inclined or nearly spherical trajectories underwent artificial circularization associated with an inadequate evolution of the Carter constant. The behavior suggests that the local least-squares problem did not reliably reproduce the secular evolution of all three orbital constants across the parameter space tested here. Because the radial-potential prescription provides substantially more stable behavior for the present application, further investigation of the least-squares approach is not a priority for the current implementation. Nevertheless, future work could examine whether improved conditioning or alternative coordinate transformations relating the angular momentum vector to the Carter constant could yield a more reliable local prescription. Identifying the conditions under which the least-squares system becomes poorly conditioned may also help determine whether such approaches could be useful in other regions of the parameter space.

\subsection{Extending the Radiation-Reaction Model}

The radiation-reaction prescription used in the present implementation is based on the leading-order, quadrupole-order fluxes derived by Peters~\citep{Peters1964}. These provide a well-established weak-field approximation and are sufficient for capturing the basic secular evolution of the orbital constants. However, the quadrupole approximation becomes increasingly limited as the orbit enters the strong-field regime, where higher-order post-Newtonian corrections and black-hole perturbation-theory calculations provide more accurate descriptions of the energy and angular-momentum fluxes \citep{Thorne1980, Nolan_et_al_2015, LISAConsort2025, Castillo_et_al_2025}.

An important advantage of the modular structure of \EMRICode{}is that the flux prescription can be replaced without fundamentally changing the underlying geodesic evolution. Higher-order post-Newtonian or perturbative fluxes could therefore be incorporated as alternative radiation-reaction models, allowing the same framework to test how increasingly sophisticated prescriptions affect the secular evolution of generic Kerr orbits. The existing Peters-based model thus provides a simple baseline against which these more accurate treatments can be compared.

A second potential extension is the inclusion of spin and higher multipole moments of the secondary body. For a non-spinning point particle, the secondary follows a Kerr geodesic to leading order in the mass ratio. A spinning secondary, however, experiences additional couplings between its spin and the background curvature, modifying its orbital evolution at higher order in the mass-ratio expansion \citep{Chicone_Mashhoon_Punsly_2005,Li_Wu_Singh_2019}. The simplest implementation could begin with spin aligned or anti-aligned with the central black-hole spin, avoiding the additional complexity associated with generic spin precession. Such an extension would provide a controlled way to investigate when secondary-spin effects become dynamically relevant and whether they can be incorporated primarily through modifications to the radiation-reaction prescription or instead require changes to the geodesic evolution itself \citep{Dolan_et_al_2014,Piovano_Maselli_Pani_2020}.

\subsection{Environmental Interactions and an Expanded Scope}

One of the principal motivations for maintaining a modular radiation-reaction framework is the possibility of extending the model beyond vacuum EMRIs. Real galactic nuclei contain dark matter cusps, stars, compact objects, gas, accretion disks, and other forms of matter that can perturb the motion of an inspiraling secondary. Over the many orbital cycles accumulated during an EMRI's life, even relatively weak perturbations can produce secular changes in the orbital evolution \citep{Copparoni_Chandramouli_Barausse_2026,Speri_et_al_2026}.

\begin{figure}[h]
\centering
\includegraphics[scale=0.6]{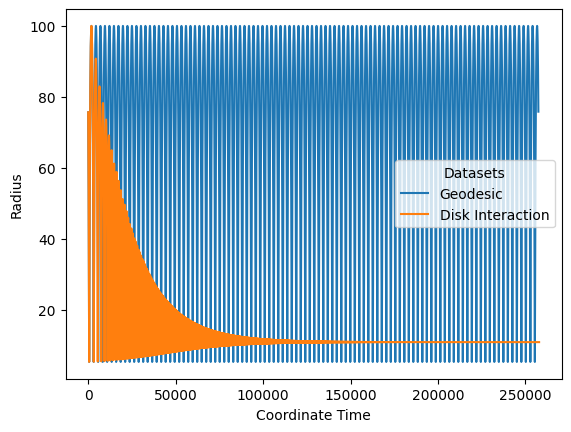}
\caption{Comparison of a toy model of repeated accretion disk interactions (orange) against a pure geodesic (blue), with initial conditions $a=0.5, p=10, e=0.9, \theta_-=\frac{\pi}{4}$. Modeling the interaction as an inelastic collision of the secondary body with a smaller mass on a circular equatorial orbit at the impact radius leads to the secondary body eventually trending towards a circular orbit.}
\label{fig:toy_disk}
\end{figure}

\begin{figure}[h]
\centering
\includegraphics[scale=0.6]{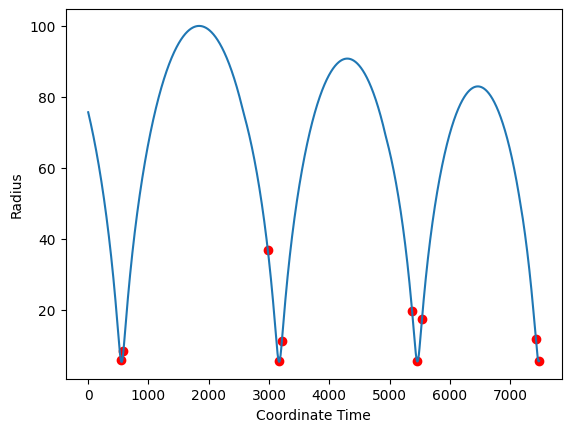}
\caption{The disk interaction toy model with disk impacts marked in red. Tracking the time and location of potential impacts could be linked to QPE phenomenology.}
\label{fig:toy_disk_impact}
\end{figure}

The most immediate environmental extension would be to include interactions between the secondary and an accretion disk or gaseous environment. At the simplest level, \EMRICode{}could identify intersections between the secondary's trajectory and a prescribed disk geometry and record the orbital phase, radius, and velocity at each encounter. This would provide a way to identify orbital configurations and encounter times that could be relevant to electromagnetic activity. Such star-disk interactions have been proposed as a possible mechanism for quasi-periodic eruptions (QPEs), particularly for inclined stellar orbits that cross an accretion disk once or twice per orbit \citep{Linial_Metzger_2023};  Recent observational and theoretical work has likewise considered disk-crossing stellar objects as a possible explanation for QPE phenomenology \citep{Chakraborty_et_al_2025}. Figures \ref{fig:toy_disk} and \ref{fig:toy_disk_impact} utilize a very simple application of the concept, in which the secondary body of the EMRI system repeatedly impacts and absorbs mass and momentum from a thin equatorial accretion disk; Fig.~\ref{fig:toy_disk} shows the hypothetical evolution of the secondary body's radial motion on an accelerated time scale, while Fig.~\ref{fig:toy_disk_impact} marks the recorded impact radii as a function of time. This 'toy model' excludes the effect of radiation reaction to more easily isolate the effect of the disk.

A natural next step would be to model the dynamical effect of those encounters. Gas drag, momentum exchange, and mass ablation could provide additional changes to the orbital energy and angular momentum, allowing disk interactions to compete directly with gravitational-wave-driven inspiral \citep{Barausse_Rezzolla_2008,Yunes_et_al_2011}. Because these effects would act as additional contributions to the evolution of the orbital constants, they could easily be incorporated into the same modular framework used for gravitational radiation reaction. The fidelity of such a model could then be increased progressively, beginning with a simple prescribed drag force and eventually incorporating a more detailed description of the disk structure and local gas dynamics.

Finally, the point-particle approximation could be relaxed to investigate partial tidal disruption and mass loss. For a stellar secondary, sufficiently close passages to the central black hole can remove some fraction of the stellar mass without completely disrupting the object \citep{Bortolas_et_al_2023}. A full treatment would require resolving the internal structure and hydrodynamics of the secondary and is therefore well beyond the scope of the present model. However, a simplified prescription could potentially estimate mass loss from quantities such as the penetration factor, approximate stellar structure, and encounter velocity, while updating the mass and orbital constants of the surviving remnant. Such a model could also provide a natural termination condition for complete disruption.

\begin{figure}[h]
\centering
\includegraphics[scale=0.6]{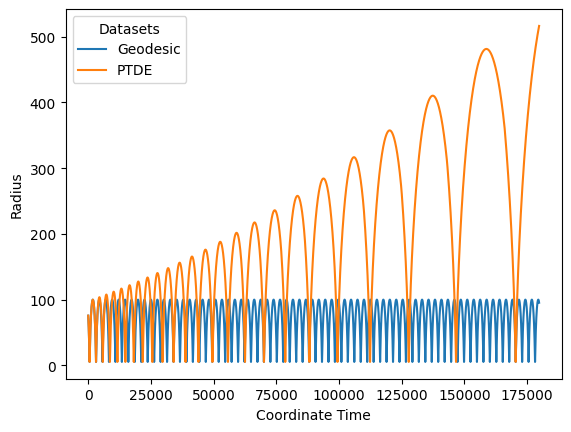}
\caption{Comparison of a toy model repeated partial tidal disruption against a pure geodesic, with initial conditions $a=0.5, p=10, e=0.9, \theta_-=\frac{\pi}{4}$. Modeling the rPTDE as a removal of 1\% of the secondary mass at each periapse, with the removed portion set on a circular orbit, leads to the remaining disrupted body taking an increasingly parabolic orbit.}
\label{fig:toy_PTDE}
\end{figure}

This extension is particularly interesting because mass loss need not affect only the luminosity or detectability of the system. Changes in the mass and internal structure of the secondary can also modify its orbital energy and angular momentum, potentially producing orbital evolution that differs qualitatively from a purely gravitational-wave-driven inspiral \citep{Payne_et_al_2021, Wang_Wang_2026}. In combination with the disk-interaction model described above, this could provide a relatively simple framework for exploring connections between EMRIs, partial tidal disruptions, and quasi-periodic eruptions without requiring a fully relativistic hydrodynamic treatment. Figure~\ref{fig:toy_PTDE} employs a similar concept to the accretion disk model, but instead {\it removes} mass and momentum from the secondary body as a function of its periapsis radius; assuming the removed portion is immediately set on a circular orbit leads to an increase in semilatus rectum and period, but a more sophisticated model could produce much more complex behavior as the system evolves. 

A related extension would be to include perturbations produced by additional massive bodies in the nuclear environment. A nearby star or stellar-mass black hole can exert a gravitational perturbation on the EMRI orbit, potentially producing resonant changes in the orbital evolution \citep{Naoz_2022}. \citet{Bonga_Yang_Hughes_2019} demonstrated that such tidal resonances provide a relativistic analogue of Kozai--Lidov-type dynamics and can imprint measurable effects on EMRI waveforms. More generally, environmental perturbations may therefore provide both a source of systematic error for vacuum EMRI models and a potential means of extracting information about the environment surrounding the central black hole. 

At higher fidelity, these perturbations could be treated not merely as additional forces acting on the secondary, but as modifications to the spacetime in which the orbit evolves. This distinction is important because a sufficiently massive or structured environment cannot necessarily be represented as a small external force while retaining an exactly Kerr background. Recent work has begun to investigate self-consistent perturbations of environmental fields around inspiraling secondaries, illustrating the more general problem that arises when the surrounding matter contributes directly to the spacetime dynamics.

\section{Conclusion}
\label{sec:conclusion}

We have presented \EMRICode, a modular numerical framework for evolving generic extreme mass-ratio inspirals in Kerr spacetime. The implementation combines adaptive numerical integration of the geodesic equations with secular evolution of the orbital constants of motion, allowing generic eccentric and inclined trajectories to be evolved over long inspiral timescales. By separating the geodesic evolution from the prescription used to modify the orbital constants, the framework also provides a flexible basis for testing and comparing different models of radiation reaction and, ultimately, additional physical effects.

A central focus of this work has been the treatment of the Carter constant, for which no direct analogue of the energy and axial angular-momentum fluxes is available within the simple approximation used here. We investigated several prescriptions for determining its evolution and found that the radial-potential method provides the most stable behavior among those considered for the parameter space explored in this work. The alternative prescriptions exhibited more significant failures, including artificial circularization and other unphysical evolution for highly inclined or nearly spherical trajectories. The radial-potential prescription nevertheless retains a systematic residual bias, particularly for nearly planar orbits, indicating that further refinement of the mapping between the local radial dynamics and the secular evolution of the Carter constant is needed.

We also compared the resulting inspirals with established approximations from the literature, including the weak-field treatment of Peters and the Kerr-based prescriptions of Glampedakis, Hughes, and Kennefick and Gair and Glampedakis. These comparisons demonstrate that \EMRICode{}reproduces the expected qualitative behavior of radiation-reaction-driven inspiral while also illustrating the differences that arise from the assumptions underlying each prescription. In particular, the comparisons provide a useful benchmark for assessing the behavior of the selected Carter-constant prescription and establish a baseline against which more sophisticated flux models can be tested.

The present implementation is intentionally a starting point rather than a complete model of astrophysical EMRIs. The modular structure of \EMRICode{}makes it possible to replace the leading-order radiation-reaction prescription with higher-order post-Newtonian or perturbative fluxes, while retaining the same underlying orbital evolution framework. Additional extensions could incorporate secondary spin, improved treatments of the Carter constant, and environmental interactions with surrounding gas or accretion disks. Relaxing the point-particle approximation could further allow the framework to explore partial tidal disruption and mass loss. These extensions would broaden the range of physical systems that can be studied while preserving the central feature of the present approach: an explicit numerical representation of the orbital dynamics that can be coupled to different models of the processes driving secular evolution.

More broadly, \EMRICode{}provides a computational framework for exploring the dynamics of EMRIs between highly idealized analytic treatments and more specialized high-fidelity calculations. Its value lies in retaining sufficient detail to follow generic Kerr trajectories while maintaining the flexibility to introduce and compare additional physical effects. This makes it a useful foundation for investigating how radiation reaction, secondary structure, and environmental interactions may combine to shape the evolution of compact objects in the strong-field environments surrounding massive black holes.

\addcontentsline{toc}{section}{References}
\bibliography{references.bib}

\end{document}